\documentclass[12pt]{article}%
\usepackage[nosort]{cite}
\usepackage[usenames, dvipsnames]{xcolor}
\usepackage{graphicx}
\usepackage{multicol}
\usepackage{amsfonts}
\usepackage{amssymb}
\usepackage{amsmath}
\usepackage{heck}
\usepackage{afterpage}
\usepackage{setspace}
\usepackage{verbatim}
\usepackage{color}
\usepackage{longtable}
\usepackage{float}
\usepackage{subcaption}
\usepackage{epsfig}
\usepackage{enumerate}
\usepackage{epstopdf}
\usepackage[enableskew, vcentermath]{youngtab}
\usepackage{adjustbox}
\usepackage{multirow}
\usepackage{tikz}
\usepackage[margin=1in]{geometry}
\usepackage{titletoc}
\usepackage[percent]{overpic}
\usepackage{gensymb}
\usepackage{mathtools}
\usepackage{tikz-cd}%
\providecommand{\U}[1]{\protect\rule{.1in}{.1in}}
\pdfoutput=1
\newsavebox{\mysavebox}

\usetikzlibrary{decorations.markings}

\numberwithin{equation}{section}

\usetikzlibrary{chains}
\allowdisplaybreaks
\tikzset{node distance=2em, ch/.style={circle,draw,on chain,inner sep=2pt},chj/.style={ch,join},every path/.style={shorten >=4pt,shorten <=4pt},line width=1pt,baseline=-1ex}

\newcommand{\ba}{\begin{eqnarray}}
\newcommand{\ea}{\end{eqnarray}}
\newcommand{\cA}{\mathcal{A}}

\newcommand{\cL}{\mathcal{L}}

\newcommand{\cO}{\mathcal{O}}

\newcommand{\Tr}{\, {\rm Tr}}

\newcommand{\be}{\begin{equation}}
\newcommand{\ee}{\end{equation}}

\tikzstyle{startstop} = [rectangle, rounded corners, minimum width=3cm, minimum height=1cm,text centered, draw=black, fill=blue!10]
\tikzstyle{startstop} = [rectangle, rounded corners, minimum width=3cm, minimum height=1cm,text centered, draw=black, fill=blue!10]
\tikzstyle{io} = [trapezium, trapezium left angle=70, trapezium right angle=110, minimum width=3cm, minimum height=1cm, text centered, draw=black, fill=blue!30]
\tikzstyle{process} = [rectangle, minimum width=3cm, minimum height=1cm, text centered, draw=black, fill=orange!30]
\tikzstyle{decision} = [diamond, minimum width=3cm, minimum height=1cm, text centered, draw=black, fill=green!30]
\tikzstyle{arrow} = [thick,->,>=stealth]
\tikzset{->-/.style={decoration={
  markings,
  mark=at position #1 with {\arrow[scale=2.4]{>}}},postaction={decorate}}}
\makeatletter \@addtoreset{equation}{section} \makeatother

\newcommand{\CFTUV}{\text{CFT}_{\text{UV}}}
\newcommand{\CFTIR}{\text{CFT}_{\text{IR}}}
\newcommand{\p}{\partial}
\renewcommand{\(}{\left(}
\renewcommand{\)}{\right)}

\newcommand{\D}{\nabla}

\colorlet{darkblue}{blue!70!black}
\colorlet{darkgreen}{green!70!black}

\usepackage[colorlinks=true,urlcolor=darkblue,linktocpage=true,linkcolor=darkblue,citecolor=darkblue]{hyperref}
\hypersetup{
colorlinks=true,
linkcolor=blue,
citecolor=magenta,
}

\begin{document}

\date{March 2021}

\title{EFT of 6D SUSY RG Flows}

\institution{PENN}{\centerline{Department of Physics and Astronomy, University of Pennsylvania, Philadelphia, PA 19104, USA}}

\institution{JHU}{\centerline{Department of Physics and Astronomy, Johns Hopkins University, Baltimore, MD 21218, USA}}

\authors{
Jonathan J. Heckman\footnote{e-mail: {\tt jheckman@sas.upenn.edu}},
Sandipan Kundu\footnote{e-mail: {\tt kundu@jhu.edu}}, and
Hao Y. Zhang \footnote{e-mail: {\tt zhangphy@sas.upenn.edu}}
}

\abstract{Motivated by its potential use in constraining the structure of 6D renormalization group flows,
we determine the low energy dilaton-axion effective field theory of conformal and global symmetry breaking
in 6D conformal field theories (CFTs). While our analysis is largely independent of supersymmetry, we
also investigate the case of 6D superconformal field theories (SCFTs), where we use the effective action
to present a streamlined proof of the 6D a-theorem for tensor branch flows, as well as to
constrain properties of Higgs branch and mixed branch flows.
An analysis of Higgs branch flows in some examples leads us to
conjecture that in 6D SCFTs, an interacting dilaton effective theory may be possible even when
certain 4-dilaton 4-derivative interaction terms vanish, because of large
momentum modifications to 4-point dilaton
scattering amplitudes. This possibility is due to the fact that in all known $D > 4$ CFTs,
the approach to a conformal fixed point involves effective strings which are becoming tensionless.}

\maketitle

\setcounter{tocdepth}{2}

\tableofcontents


\newpage

\section{Introduction \label{sec:INTRO}}

One of the important features of quantum field theories is that they can be
organized according to a hierarchy of scales which can be related by
renormalization group (RG) flows. Conformal field theories (CFTs) are of particular
significance since they arise from fixed points of the renormalization group,
and the general structure of these theories indicates that there is a sense in
which information is lost in passing from the ultraviolet (UV)\ to the
infrared (IR).

This intuition can be sharpened significantly in the context of conformal
field theories (CFTs)\ in even spacetime dimensions and the structure of its
conformal anomalies, namely the behavior of the trace of the stress energy
tensor in curved backgrounds. Starting from a conformal fixed point, a
perturbation will drive a flow to another theory in the IR. Reference \cite{Zamolodchikov:1986gt}
established that for 2D CFTs, there exists a $c$-function which decreases monotonically
along an RG\ flow, and so in particular $c_{\text{UV}}>c_{\text{IR}}$ for the
central charges of a UV and IR\ CFT connected by such a flow. In reference
\cite{Cardy:1988cwa} it was conjectured that a similar c-theorem should hold for
4D\ theories for the Euler anomaly $a$. Many intermediate results such as
\cite{Schwimmer:2010za, Anselmi:1997am,Intriligator:2003jj} provided strong evidence that for 4D\ RG flows,
$a_{\text{UV}}>a_{\text{IR}}$. In reference \cite{Komargodski:2011vj} a convincing
demonstration of the 4D\ $a$-theorem was presented based on the structure of
dilaton scattering in such theories. It is widely expected that
there should be an $a$-theorem for 6D CFTs, though a straightforward extension
of 4D\ dilaton scattering methods to six
dimensions runs into numerous obstacles \cite{Elvang:2012st,Kundu:2019zsl}.

Indeed, on general grounds, 6D CFTs are more
difficult to characterize because even defining them requires passing far
beyond perturbations of a Gaussian fixed point. The first examples of
interacting fixed points in six dimensions were only found using methods from
string theory \cite{Witten:1995zh,Strominger:1995ac}, and were recognized as such based on the
strong arguments of reference \cite{Seiberg:1996qx}. As of this writing, the only known
examples are supersymmetric. This singles out six spacetime
dimensions as especially important because the classification results of
reference \cite{Nahm:1977tg} also show that six is the maximum spacetime dimension which
can support superconformal symmetry. Recently, significant progress has been made in classifying
6D SCFTs via the geometry of F-theory backgrounds
\cite{Heckman:2013pva, Heckman:2015bfa} (see also \cite{Bhardwaj:2015xxa, Bhardwaj:2019hhd} as well as
the review \cite{Heckman:2018jxk}). In particular, a number of analyses of RG flows in deformations of
6D SCFTs have now been performed, see e.g. \cite{Heckman:2015ola, Cordova:2015vwa, Cordova:2015fha, Heckman:2015axa, Louis:2015mka, Cordova:2016xhm, Buican:2016hpb, Cordova:2016emh, Heckman:2016ssk, Heckman:2018pqx}.

Motivated by the prospect of determining possible constraints on the structure of 6D RG flows,
our aim in this paper will be to construct the dilaton-axion effective action associated with
spontaneous breaking of conformal symmetry, when it is accompanied by the breaking of a global continuous symmetry.
This sort of situation is actually rather generic in 6D SCFTs, and arises whenever the R-symmetry is broken.
We comment that in all known examples where the R-symmetry is broken, there are additional
axions and scalars generated from the breaking of flavor symmetries which fill out hypermultiplets
on the Higgs branch moduli space. That being said, we shall mainly
focus on the most explicit case where the global symmetry contains
an $SU(2)$ factor, leaving the other contributions implicit.

Similar to the 4D case \cite{Kundu:2020bdn}, we find that the dilaton-axion
effective action $S_{\text{eff}}[\phi,\xi_a]$, up to 6-fields and 6-derivative,
admits a decomposition
\be\label{intro:action}
S_{\rm eff}[\phi,\xi]=\int d^6x\left(-\frac{1}{2}\( (\p \phi)^2+ (\p \xi_a \cdot \p \xi_a)\)+\cL_{\rm dilaton}[\phi]+\cL_{\rm axion}[\xi]+\cL_{\rm mixed}[\phi,\xi]\right)\ ,
\ee
where, $\cL_{\rm dilaton}[\phi]$ is the part of the effective action which is independent of the global symmetry breaking. Hence, $\cL_{\rm dilaton}[\phi]$ captures the interactions of the dilaton effective action of \cite{Elvang:2012st}. Likewise, $\cL_{\rm axion}[\xi]$ captures the axion interactions that result from the global symmetry breaking. Moreover, there can be mixed interactions $\cL_{\rm mixed}[\phi,\xi]$ containing at least four derivatives, two dilatons, and two axions. An important consequence of the above decomposition is that the global symmetry breaking will not affect any proof of the 6D $a$-theorem involving only $\cL_{\rm dilaton}[\phi]$.

Additional information can be extracted when we assume the existence of supersymmetry.
In perturbations of 6D superconformal fixed points, no marginal or relevant
deformations that preserve supersymmetry are possible, and instead all flows are triggered by vacuum
expectation values (VEVs) for operators \cite{Louis:2015mka, Cordova:2016xhm}.
There are essentially three ways in which a
supersymmetry preserving flow can be triggered, based on the classification
results of reference \cite{Cordova:2016xhm} which are known as tensor branch, Higgs
branch, and mixed branch flows. As the names suggest, in a tensor branch flow
the scalar component of a 6D\ tensor multiplet attains a non-zero value, while
in a Higgs branch flow, a combination of scalars break the $SU(2)$ R-symmetry
of the system. A mixed branch flow amounts to a combination of tensor and
Higgs branch flows.

One of the goals of this paper is to explore the structure of the dilaton-axion
effective theory for 6D mixed branch flows. Our result implies some non-trivial consequences for the structure of
dilaton scattering in the corresponding effective field theory. In particular,
we see that in a mixed branch flow where we have a combination of tensor
branch and Higgs branch deformations depending on the respective scales $f_t$ and $f_H$,
the respective contributions to
the term $\phi^2 \Box^2 \phi^2$ have an interesting dependence on the ratio $f_t / f_H$.
In particular, this strongly suggests the existence of a smooth limit in which we
either switch off the Higgs branch deformation or the tensor branch deformation completely.

An important element of this analysis is that it permits us to access some aspects of Higgs branch flows via the more
general case of mixed branch flows. Indeed, many 6D SCFTs have a deformation to a tensor branch in which the theory on the tensor branch
\textit{also} has a Higgs branch. In this case, we can explicitly track in the resulting low energy gauge theory how various contributions to the dilaton-axion effective action are actually generated. In contrast to the case of 4D CFTs, we present some explicit examples where all contributions to the $\phi^2 \Box^2 \phi^2$ interaction terms involve a suppression scale with at least two non-trivial powers of $f_t$. 
Said differently, we find no way to generate a purely Higgs branch contribution to this interaction term.

This is a rather surprising result, and it leads us to conjecture that for Higgs branch flows of 6D SCFTs,
the term $\phi^2 \Box^2 \phi^2$ actually vanishes. The reason this does not immediately contradict the well-known constraint
of reference \cite{Adams:2006sv} is that to actually extract a dispersion relation, one must make
some specific (and usually very well-motivated) assumptions about the large momentum behavior of dilaton 4-point scattering amplitudes. This is problematic in six dimensions, because all known examples involve effective strings which have tension which is tuned to zero at the conformal fixed point. In the case of tensor branch flows, this is not much of an issue because the effective field theory presupposes that the energy scale is less than $f_t$, and similar considerations apply for mixed branch flows. The situation is far more subtle for Higgs branch flows,
since in this case $f_t = 0$, namely the effective strings on the tensor branch are becoming tensionless. Again, this appears to be a feature unique to $D > 4$ CFTs, since in the case of $D \leq 4$ CFTs, the approach to a conformal fixed point only involves particles becoming massless.

The rest of this paper is organized as follows. In section \ref{sec:EFT} we set up the
general formalism of a dilaton-axion effective action for 6D\ CFTs in which conformal and a global symmetry
are both spontaneously broken. In section \ref{sec:SUSY} we
specialize this effective action to the case of 6D SCFTs
and show how supersymmetry leads to additional constraints.
Section \ref{sec:example} studies some examples of 6D SCFTs with mixed branch
flows and the consequences for Higgs branch flows.
These examples motivate the statements of section \ref{sec:CONJ} where
we  present a conjecture on the vanishing of the term $\phi^2 \Box^2 \phi^2$
in Higgs branch flows. Section \ref{sec:CONC} contains our conclusions.
Some additional aspects of the dilaton-axion effective
action are included in Appendix \ref{app:DILAX}.

\section{The Dilaton-Axion Effective Theory \label{sec:EFT}}

In this section we find the general dilaton effective action associated with the breaking of conformal symmetry in six dimensions. In
the situations of interest, especially Higgs branch and mixed branch flows, this is often accompanied by spontaneous breaking of a global $SU(2)$ symmetry. So, our aim will be to write down an effective action for the associated dilaton-axion effective theory.

Now, there is a crucial difference between spontaneous breaking of global symmetries and spacetime symmetries. The Goldstone modes that are generated as a consequence of spacetime symmetry breaking are not all independent \cite{Volkov:1973vd}. In particular, when conformal symmetry is spontaneously broken to the Poincar\'e symmetry, there is only one independent Goldstone mode -- the dilaton $\tau$. The rest of the Goldstone modes $a_\mu$  generated from the broken special conformal generators are constrained by the inverse Higgs effect $a_\mu \sim \partial_\mu e^{\tau}$ \cite{Salam:1970qk,Isham:1970xz,Isham:1970gz,Ivanov:1975zq,Low:2001bw,McArthur:2010zm,Hinterbichler:2012mv}. This is reminiscent of the constraint on the vector field  of a 6D $(1,0)$ linear multiplet (see for example \cite{Bergshoeff:1985mz}).

In addition, the spontaneously broken UV $SU(2)$ symmetry generates three massless pseudo-scalars $\beta_a$, $a=1,2,3$. Hence, in the IR we end up with $\CFTIR$, the dilaton $\tau$,  and axions $\beta_a$
 \begin{align}\label{IR}
&\CFTUV   \nonumber\\
&~~\Downarrow \\
&\CFTIR  + S_{\rm eff}[\tau,\beta_a]\ .\nonumber
 \end{align}
We also assume that an $SU(2)$ global symmetry gets restored in $\CFTIR$. This requirement stems from the fact that eventually we will specialize to the supersymmetric case where the global $SU(2)$ symmetry is the $R$-symmetry of $(1,0)$ SCFTs. However, the discussion of this section is more general and applies to non-supersymmetric 6D flows  as well. In the general case, we  assume that $\CFTIR$ is invariant under the UV $SU(2)$ transformations.

The effective action $S_{\rm eff}[\tau,\beta_a]$ can be derived by simply extending the analysis of \cite{Kundu:2020bdn} to 6D. First, we couple the theory to a background metric $g_{\mu\nu}(x)$ and a background $SU(2)$ gauge potential
\be
A_\mu= A^a_\mu(x) \sigma_a\ ,
\ee
where $\sigma_a$, $a=1,2,3$ are the Pauli matrices.\footnote{We are using the notation in which $\sigma_a \sigma_b = \delta_{ab}\mathbb{I}+i \varepsilon_{abc}\sigma_c$.} In the presence of these background fields, the 6D conformal trace anomaly is given by \cite{Deser:1976yx,Duff:1977ay,Fradkin:1983tg,Deser:1993yx}
\be\label{anomaly0}
 \langle T^\mu_\mu\rangle = a E_6+ \sum_{i=1}^3 c^{(i)} I_i+ I_F\
 \ee
 up to total derivative terms. Total derivative terms are not important since they can be removed by adding counterterms in the UV theory. In the above expression, $E_6$ is the 6D Euler density
\be
 E_6= \frac{1}{8}\delta^{\mu_1 \mu_2 \mu_3 \mu_4 \mu_5 \mu_6}_{\nu_1 \nu_2 \nu_3 \nu_4 \nu_5 \nu_6}R^{\nu_1\nu_2}_{~~~~\mu_1 \mu_2}R^{\nu_3\nu_4}_{~~~~\mu_3 \mu_4}R^{\nu_5\nu_6}_{~~~~\mu_5 \mu_6}
 \ee
 and $a$ is the corresponding Euler anomaly. Similarly, the central charges $\{c^{(i)}\}$ are associated with conformal invariants:
 \begin{align}
 I_1=& W_{\gamma \alpha \beta \delta}W^{\alpha \mu \nu \beta}W_{\mu~~\nu}^{~\gamma \delta}\ , \notag\\
 I_2=& W_{\alpha \beta}^{~~\gamma \delta}W_{\gamma \delta}^{~~\mu \nu}W_{\mu \nu}^{~~\alpha \beta}\ , \notag\\
 I_3=& W_{\alpha \gamma \delta \mu}\left(\nabla^2 \delta^\alpha_\beta+4 R^\alpha_\beta-\frac{6}{5}R \delta^\alpha_\beta \right)W^{\beta \gamma \delta \mu}\ ,
 \end{align}
where, $W_{\mu\nu\alpha\beta}$  is the Weyl tensor. For supersymmetric theories, the $\{c^{(i)}\}$ are not all independent.
In the relation (\ref{anomaly0}), $I_F$ represents terms due to the background gauge field \cite{Cordova:2019wns}
\be
I_F=f_1 \mathrm{Tr} \left({\mathcal D}_\mu F^{\mu\lambda} {\mathcal D}^\nu F_{\nu\lambda}\right)+f_2 W_{\mu\nu\alpha\beta}\mathrm{Tr}\left(F^{\mu\nu}F^{\alpha\beta}\right)+f_3 \mathrm{Tr}\left(F^{\mu\nu}F_{\mu\alpha}F^{\alpha}_{\ \nu}\right)+\cdots\ ,
\ee
where $F$ is the field strength for the background gauge field.\footnote{Note that $\nabla_\mu$ is the spin connection covariant derivative, whereas ${\mathcal D}_\mu$ is the combined spin and gauge connection covariant derivative.} The dots represent terms that are either total derivatives or terms that are completely fixed by conformal invariance. Terms that are completely fixed by conformal invariance cannot change under RG flows and hence will not play any role in our discussion. The coefficients $f_1,f_2,f_3$ are arbitrary, however, for supersymmetric theories they are related to $c_1,c_2$ and $c_3$ \cite{Cordova:2019wns}.  The current $j_\mu$ associated with the $SU(2)$ symmetry can also be anomalous $\langle {\mathcal D}_\mu j^\mu \rangle \neq0$ in the presence of background fields. We do not need to know the exact form of the anomaly, since, $\sqrt{g} \langle {\mathcal D}_\mu j^\mu\rangle$ is both gauge and Weyl invariant. In fact, in the flat space limit with no gauge field, only the Euler anomaly term will contribute to the effective action.

We follow \cite{Kundu:2020bdn} to derive $S_{\rm eff}[\tau,\beta_a]$ by studying the variation of the action under diff$\times$Weyl transformations and gauge transformations. Under Weyl transformations
 \be\label{weyl}
 g_{\mu\nu}(x)\rightarrow e^{2s(x)}g_{\mu\nu}(x)\ , \qquad \tau(x)\rightarrow \tau(x)+s(x)\ .
 \ee
Gauge transformations act on the gauge field in the usual way
 \begin{align}
 A_\mu(x) \rightarrow \Omega(x)  A_\mu(x) \Omega^{-1}(x)+i \Omega(x) \nabla_\mu \Omega^{-1}(x)\ ,
 \end{align}
 where $\Omega(x)=e^{i \pi_a(x) \sigma_a} \in SU(2)$, and the axions $\beta$ transform as
 \be
 \beta(x)\rightarrow\beta(x)+ \pi(x)\ .
 \ee

We now invoke the anomaly matching arguments of \cite{Schwimmer:2010za,Komargodski:2011vj} which imply that the changes in anomalies from the UV to the IR must be compensated by the Goldstone bosons. This imposes (at the linearized order)
\be\label{variation1}
\text{Weyl:} \,\,\, \delta_s S_{\rm eff}[g_{\mu\nu},A_\mu;\tau,\beta]=\int d^6x \sqrt{-g} s(x)\left(\Delta a E_6+ \sum_{i=1}^3 \Delta c^{(i)} I_i +\Delta I_F\right)
\ee
and
\be\label{variation2}
\text{Gauge:} \,\,\, \delta_\pi S_{\rm eff}[g_{\mu\nu},A_\mu;\tau,\beta]=\int d^6x \sqrt{-g} \ \mathrm{Tr}\left( \pi(x) \Delta \langle{\mathcal D}_\mu j^\mu\rangle \right)\ ,
\ee
where $\Delta (\cdots)$ represents the change of a quantity under the RG flow (for example, $\Delta a=a_{\rm UV}-a_{\rm IR}$). Of course, all IR anomalies must be understood as the total anomalies of $\CFTIR$ and all the Goldstone bosons.

The variational equations (\ref{variation1}) and (\ref{variation2}) can be solved systematically to obtain $S_{\rm eff}$. We follow \cite{Kundu:2019zsl,Kundu:2020bdn} and decompose the effective action in the following way
\begin{align}\label{simple}
S_{\rm eff}[g_{\mu\nu},A_\mu;\tau,\beta]&=\int d^6x \sqrt{-g}\ \mathrm{Tr}\left( \beta(x) \Delta \langle{\mathcal D}_\mu j^\mu\rangle\right)\nonumber\\*
&+\int d^6x \sqrt{-g} \tau(x)\left(\Delta a E_6+ \sum_{i=1}^3 \Delta c^{(i)} I_i +\Delta I_F\right)+\delta S_{\rm WZ}+S_{\rm inv}\ .
\end{align}
The first term in the above equation is designed to simply generate the correct gauge variation (\ref{variation2}).  Since the Euler density $E_6$ is not Weyl invariant, it is more complicated to generate the Weyl variation (\ref{variation1}). For example, the second term in equation (\ref{simple})  generates the correct Weyl variation (\ref{variation1}) plus an extra term $\Delta a\int d^6x \sqrt{-g} \tau(x)\delta_s E_6$. This extra term is cancelled by adding a correction term $\delta S_{\rm WZ}$. In addition, we can always add terms which are invariant under gauge and Weyl transformations; $S_{\rm inv}$ represents all such invariant terms that are allowed by symmetry.

The effective action simplifies greatly when we take the flat space limit without the background gauge field. This mainly comes from the fact that $\delta S_{\rm WZ}$ is uniquely fixed by $\Delta a$  up to terms that are invariant under both diff$\times$Weyl  transformations and gauge transformations. Moreover, in the flat space limit without the background gauge field, $\delta S_{\rm WZ}$ simplifies further \cite{Elvang:2012st}
\be\label{uni}
\delta S_{\rm WZ}|_{g_{\mu\nu}=\eta_{\mu\nu},A_\mu=0}=3\Delta a\int d^6x \tau \Box^3 \tau + \cdots\ ,
\ee
where, the dots represent terms that can be absorbed in $S_{\rm inv}$. We now take the flat space limit without the background gauge field of  the effective action (\ref{simple}). In this limit  only $\delta S_{\rm WZ}$ and $S_{\rm inv}$ contribute
 \be
 S_{\rm eff}[\tau,\beta]=3\Delta a\int d^6x\ \tau \Box^3 \tau + S_{\rm inv}[\tau,\beta]\ .
 \ee
Hence, the effective action is completely fixed by symmetries. The invariant part of the action can be efficiently constructed by defining Weyl invariant and gauge covariant combinations \cite{Kundu:2020bdn}
\be\label{metric0}
\hat{g}_{\mu\nu}(x)=e^{-2\tau(x)}g_{\mu\nu}(x)\ , \qquad \hat{A}_\mu(x)=A_\mu(x) - i e^{i \beta(x)} \p_\mu e^{-i \beta(x)} \ .
\ee
Of course, $S_{\rm inv}$ constructed only from $\hat{g}_{\mu\nu}$ and $\hat{A}_\mu$ will miss Wess-Zumino type terms that shift by total derivatives under gauge and Weyl transformations. But such terms will not contribute to the effective action containing up to six fields. Up to six derivatives, the most general $S_{\rm inv}[\hat{g}_{\mu\nu},\hat{A}_\mu]$ can be written as
\begin{align}\label{inv}
S_{\rm inv}[\hat{g}_{\mu\nu},\hat{A}_\mu]=-\frac{f^4}{2}\int d^6x \sqrt{-\hat{g}}&\left( \frac{\hat{R}}{5}+2\gamma_0^2 \hat{g}^{\mu\nu} \mathrm{Tr}\left(\hat{A}_\mu\hat{A}_\nu\right)\right) \nonumber\\
&+\int d^6x \sqrt{-\hat{g}}\left(\cL_{\rm conf}[\hat{g}_{\mu\nu}]+\cL_{SU(2)}[\hat{g}_{\mu\nu},\hat{A}_\mu]\right)\ ,
\end{align}
where the ``decay constant'' $f$ has dimension of mass and $\gamma_0$ is a real dimensionless coefficient. Note that $\cL_{\rm conf}[\hat{g}_{\mu\nu}]$ contains four and six derivative invariants that are constructed only out of the metric but not the gauge field.
By contrast, $\cL_{SU(2)}[\hat{g}_{\mu\nu},\hat{A}_\mu]$ contains four and six derivative invariants with at least one $\hat{A}_\mu$ field. The advantage of this decomposition is that $\cL_{SU(2)}[\hat{g}_{\mu\nu},\hat{A}_\mu]=0$ for $\beta_a=0$ in the flat space limit with no background gauge field. Putting everything together, $S_{\text{eff}}[\tau,\beta]$ is given by:\footnote{Our convention is that the spacetime metric is mostly $+$'s.}
\begin{align}\label{eff2}
S_{\text{eff}}[\tau,\beta]=& \int d^6x \(-2f^4e^{-4\tau}\left(\(\p \tau\)^2+\frac{1}{2}\gamma_0^2 \Tr\(\p_\mu e^{i \beta} \p^\mu e^{-i \beta}\)\)+3\Delta a \tau \Box^3 \tau\)\nonumber \\
&+\int d^6x\ e^{-6\tau}\(\cL_{\rm conf}[\hat{g}_{\mu\nu}]+\cL_{SU(2)}[\hat{g}_{\mu\nu},\hat{A}_\mu]\)_{g_{\mu\nu}=\eta_{\mu\nu}, A_\mu=0}+\cdots\ ,
\end{align}
where the dots represent higher order terms. The equations of motion at the two derivative level are given by
\be
\Box \tau=2\(\p \tau\)^2-\gamma_0^2  \Tr\(\p_\mu e^{i \beta} \p^\mu e^{-i \beta}\)\ , \qquad \Box \beta=4\(\p \tau \cdot \p \beta\)+\cO(\beta^3,\p^2)\ .
\ee
It should be noted that terms that vanish once we impose the on-shell condition for the dilaton and the axion can be removed by performing field redefinitions. However, this will generate interaction terms at higher orders.

We now focus on the invariant terms $\cL_{\rm conf}[\hat{g}_{\mu\nu}]$ that are constructed only from the Weyl-invariant metric (\ref{metric0}). At each order in the derivative expansion, there are a finite number of invariants. In particular, following \cite{Elvang:2012st} we can write
\be
\cL_{\rm conf}[\hat{g}_{\mu\nu}]=-\frac{\hat{b} f^2}{2}\hat{R}^{\mu\nu}\hat{R}_{\mu\nu}+ \frac{b'}{100} f^2 \hat{R}^2+b_1\hat{R}^3+b_2 \hat{R}\hat{R}^{\mu\nu}\hat{R}_{\mu\nu}+b_3 \hat{R}\hat{\Box}\hat{R}+\cO(\p^8)\ ,
\ee
 where all $b$-coefficients are dimensionless and numerical factors are chosen for later convenience. Of course, both $\hat{R}$ and $\hat{R}_{\mu\nu}$ are constructed using the Weyl-invariant metric (\ref{metric0}).  In the flat space limit with no gauge field, this action can be further simplified by utilizing the equation of motion of $\tau$ appropriately. For example, the Ricci scalar $\hat{R} \sim \Box \tau - 2\(\p \tau\)^2$ can be rewritten in terms of only axions by using the $\tau$ equation of motion. For our purpose, it is sufficient to keep only all-dilaton interactions. So, we can simplify
 \be
 \int d^6x \ e^{-6\tau}\cL_{\rm conf}[\hat{g}_{\mu\nu}]=4\hat{b} f^2\int d^6x\ e^{-\tau}\Box^2 e^{-\tau} +\cdots \ ,
 \ee
 where dots represent interactions involving axions that can be absorbed in $\cL_{SU(2)}[\hat{g}_{\mu\nu},\hat{A}_\mu]$.

\subsection{Effective Action for Propagating Modes}

Our eventual aim is to study dilaton scattering amplitudes. To accomplish this, we need to have a
convenient way to isolate the physical degrees of freedom which define on shell asymptotic scattering states.
With this in mind, we first perform field redefinitions so that we have canonically normalized kinetic terms
for our dilaton and axions. We refer to these as the physical dilaton and physical axion in what follows.

Along these lines, note that:
\be\label{eq:kinetic}
\frac{1}{2} \Tr\(\p_\mu e^{i \beta} \p^\mu e^{-i \beta}\)=  \(\p \beta_a\cdot \p \beta_a\)+\frac{1}{6}\epsilon_{abc}\epsilon_{ab'c'}\p_\mu\(\beta_{b}\beta_{b'}\) \p^\mu\( \beta_{c}\beta_{c'}\)+\cO(\p^2, \beta^6)\ .
\ee
Thus we perform a field redefinition
\be
e^{- 2\tau } e^{-2 i \gamma_0 \beta_a \sigma_a}= {\mathbf 1}\(1-\frac{\phi}{f^2}\) - \frac{i}{f^2} \sigma_a \xi_a
\ee
where, $\phi$ is the {\it physical dilaton} and $\xi=\xi_a \sigma_a$ is the {\it physical axion}. This implies
\be
\beta_a=\frac{\xi_a }{2\gamma_0 f^2} \(1+\frac{\phi }{f^2}+ \frac{3\phi^2-\xi^2}{3 f^4}\cdots \) \ , \qquad     \tau= \frac{\phi}{2f^2}+\frac{\phi^2-\xi^2}{4f^4}+\frac{\phi^3-3\phi \xi^2}{6 f^6}+\cdots\ ,
\ee
where $\xi^2\equiv \xi_a \xi_a$.

Finally, we can write the low energy dilaton-axion effective action (\ref{eff2}) in the following form
\begin{align}\label{eq:eff}
S_{\rm eff}[\phi,\xi]=\int d^6x\left(-\frac{1}{2}\( (\p \phi)^2+ (\p \xi_a \cdot \p \xi_a)\)+\cL_{\rm dilaton}[\phi]+\cL_{\rm axion}[\xi]+\cL_{\rm mixed}[\phi,\xi]\right) \ ,
\end{align}
where, $\cL_{\rm dilaton}[\phi]$ is the dilaton interactions \cite{Elvang:2012st} (see also \cite{Kundu:2019zsl})
\begin{align}\label{eq:dilaton}
\cL_{\rm dilaton}[\phi]=&\frac{\hat{b}}{2f^4}\phi^2 \Box^2 \phi+\frac{3\Delta a}{4f^6}\phi^2 \Box^3 \phi\\
+&\frac{\hat{b}}{f^6}\left( \frac{1}{4}\phi^3 \Box^2 \phi +\frac{1}{16} \phi^2 \Box^2 \phi^2\right)+\frac{\Delta a}{f^8}\left( \frac{1}{2}\phi^3 \Box^3 \phi +\frac{3}{16} \phi^2 \Box^3 \phi^2\right)\nonumber\\
+&\frac{\hat{b}}{32f^8}\left(5 \phi^4 \Box^2 \phi +2 \phi^3 \Box^2 \phi^2\right)+\frac{\Delta a}{8f^{10}}\left( 3\phi^4 \Box^3 \phi +2 \phi^3 \Box^3 \phi^2\right)\nonumber \\
+&\frac{\hat{b}}{128f^{10}}\left(14 \phi^5 \Box^2 \phi+5 \phi^4 \Box^2 \phi^2+2 \phi^3 \Box^2 \phi^3\right)\nonumber\\
+&\frac{\Delta a}{240f^{12}}\left( 72 \phi^5 \Box^3 \phi+45 \phi^4 \Box^3 \phi^2+20 \phi^3 \Box^3 \phi^3\right)\  \nonumber\\
+&\cO\(\phi^7\)\nonumber
\end{align}
up to total derivatives that do not contribute to amplitudes. Observe that part of the dilaton effective action is completely fixed by the change in the Euler anomaly $\Delta a=a_{\rm UV}-a_{\rm IR}$.  The dimensionless coefficient $\hat{b}$, in general, depends on the details of the symmetry breaking. However, it satisfies a positivity condition
\be
\hat{b} \ge 0
\ee
which follows from unitarity/causality \cite{Elvang:2012st,Kundu:2019zsl}. The crucial point for us is that the \textit{same} parameter $\hat{b}$ shows up in several dilaton interaction terms. Moreover, for supersymmetric flows, there are massless fermionic degrees of freedom that we will ignore in the rest of the paper.

Likewise, the interactions $\cL_{\rm mixed}[\phi,\xi]$ contain at least four derivatives, two dilatons, and two axions. On the other hand, the axion interactions of $\cL_{\rm axion}[\xi]$ have four or more fields with two or more derivatives. The two-derivative interactions in $\cL_{\rm axion}[\xi]$ come from the second term of (\ref{eq:kinetic}). Two-derivative axion interactions, if present, lead to radiative corrections. The exact form of these interactions will not be important for our purpose. However, for the sake of completeness we transcribe them in Appendix \ref{app:DILAX}.

Supersymmetry introduces additional constraints on the structure of the effective action. That being said, our result applies
more broadly and will likely be important in establishing the 6D a-theorem in both the supersymmetric and non-supersymmetric settings.

There is an additional advantage of this effective field theory formalism. In supersymmetric theories, the spontaneous breaking of conformal symmetry is often  accompanied by breaking of not only the $R$-symmetry but also other global symmetries. One can easily repeat the argument of this section to conclude that breaking of additional global symmetries will not affect $\cL_{\rm dilaton}[\phi]$ at the 6-field 6-derivative order (see appendix \ref{app:DILAX}). Hence, breaking of other global symmetries will not interfere with any proof of the $a$-theorem obtained from $\cL_{\rm dilaton}[\phi]$. This is a general feature of the dilaton-axion effective theory which is true even in 4D \cite{Kundu:2020bdn}.

\subsection{Amplitudes}
From the dilaton-axion effective action it is clear that the dilaton amplitudes, up to order $\cO(p^6)$, are completely independent of any global symmetry breaking. In particular, the 4-point dilaton scattering amplitude is still given by \cite{Elvang:2012st}
\be\label{amp:dilaton}
\cA(\phi\phi\phi\phi)= \frac{\hat{b}}{2f^6}\(s^2+t^2+u^2\)+\frac{9}{2f^8}\(\Delta a-\frac{2}{3}\hat{b}^2\)stu+\cO\(\frac{1}{f^{10}}\)\ ,
\ee
where, $s,t,u$ are the usual Mandelstam variable. Similarly, 5-pt and 6-pt dilaton amplitudes are also given by reference \cite{Elvang:2012st}.

Amplitudes involving axions depend on the details of the global symmetry breaking. Specifically for $G=SU(2)$, the dilaton-axion 4-point amplitude can be obtained from (\ref{app:mixed})
\be
\cA(\phi\phi\xi_a\xi_b)=\frac{2\delta_{ab}}{f^6}\(2\tilde{B}_2 s^2 +\tilde{B}_1 t^2+ \tilde{B}_1 u^2 \)+\cO\(\frac{1}{f^{8}}\)\ .
\ee
In the above, we have made reference to the coefficients $\tilde{B}_i$ which appear in the dilaton-axion effective action of Appendix \ref{app:DILAX}. Axion 4-point amplitudes are more complicated because of the 2-derivative interaction in (\ref{app:axions}). In particular, for $G=SU(2)$ because of this 2-derivative interaction
\be
\cA(\xi_a\xi_a\xi_b\xi_b)=\frac{12B_0}{f^4}s +\frac{8B_1}{f^6}s^2+\frac{4B_2}{f^6}(t^2+u^2) +\cO\(\frac{1}{f^{8}}\) \qquad \text{(no sum)}
\ee
for $a\neq b$. Note that the 1-loop radiative contribution from the $B_0$ term contributes at sub-sub-leading order $1/f^8$. In contrast, we observe that in 4D, the radiative corrections contribute at sub-leading order \cite{Kundu:2020bdn}.

However, note that any 4-point amplitude of identical axions does not have an $s$ term:
\be\label{ampl:axion}
\cA(\xi_a\xi_a\xi_a\xi_a)=\frac{8B_1}{f^6}\(s^2+t^2+u^2\) + \cO\(\frac{1}{f^{8}}\) \qquad \text{(no sum)}
\ee
which also gets a radiative correction from the $B_0$ term only at the order $1/f^8$. The general structure of the axion amplitudes are the same for any non-abelian symmetry group $G$.\footnote{For $G=U(1)$ there is no 2-derivative 4-axion interaction and the amplitude is precisely (\ref{ampl:axion}).}

\section{6D Supersymmetric Flows \label{sec:SUSY}}

Having laid out the general structure of the dilaton-axion effective action, we now specialize further to RG flows which preserve $\mathcal{N} = (1,0)$ supersymmetry. In this context, we identify the global $SU(2)$ with the R-symmetry of a 6D SCFT. We assume that Poincar\'{e} supersymmetry is preserved. All known $(1,0)$ SCFTs have moduli spaces of vacua that break conformal symmetry spontaneously. In fact, a general result on supersymmetry preserving deformations of SCFTs is that there are no relevant or even marginal operator deformations \cite{Louis:2015mka, Cordova:2016xhm,Cordova:2016emh}. Rather, all flows are triggered by VEVs of operators. There are exactly three types of 6D RG moduli space flows that preserve $\mathcal{N} = (1,0)$, given by tensor branch, Higgs branch and mixed branch flows.
We now discuss aspects of each of these in turn.

\subsection{Tensor Branch Flows and the $a$-Theorem}
Let us first revisit the case of tensor branch flows, where a proof of the $a$-theorem was given
in reference \cite{Cordova:2015fha}. This is the branch of the moduli space parameterized by the VEVs of tensor multiplet scalars.  Now, in these moduli space flows, the $SU(2)_R$ symmetry is
unbroken on the tensor branch, namely the dilaton is the scalar component of some linear combination of tensor multiplet scalars.
Because no global zero-form symmetries are broken, we need not discuss the axions, and the low energy effective action only contains the dilaton
\be\label{action:tensor}
S_{\rm eff}[\phi]=\int d^6x\(-\frac{1}{2} (\p \phi)^2+\cL_{\rm dilaton}[\phi]\right)\ ,
\ee
where, $\cL_{\rm dilaton}[\phi]$ is given by (\ref{eq:dilaton}). This effective action can be further simplified by removing the 3-pt interactions in (\ref{eq:dilaton}) by a field redefinition. This leads to a simplified but equivalent effective action that describes  4-point, 5-point, and 6-point amplitudes up to orders $p^6$
\be\label{final}
S_{\text{eff}}[\phi]=\int d^6x \left(-\frac{1}{2}(\partial \phi)^2+\mathcal{L}_{\phi^4}+\mathcal{L}_{\phi^5}+\mathcal{L}_{\phi^6}+\cdots \right)\ ,
\ee
where $ \mathcal{L}_{\phi^n}$ represents effective $n$-pt interaction of $\phi$ fields.
The effective interactions which contribute to 4-point amplitudes up to orders $p^6$ are given by
\begin{align}
\mathcal{L}_{\phi^4}=&\frac{\hat{b}}{16f^6} \phi^2 \Box^2 \phi^2+\frac{3}{16f^8}\left(\Delta a-\frac{2}{3}\hat{b}^2\) \phi^2 \Box^3 \phi^2+\cdots \ ,\label{int4}
\end{align}
One can easily check that the above 4-point interactions produce the same 4-point amplitude (\ref{amp:dilaton}) as the effective action (\ref{eq:dilaton}). For tensor branch flows, the $a$-theorem can be proved directly from the above effective action (\ref{int4}).  The argument is similar to that of \cite{Cordova:2015fha}, but not exactly the same, a point we now elaborate on.

Supersymmetry imposes additional constraints on the dilaton-axion effective action. Utilizing superconformal representation theory \cite{Cordova:2016xhm,Cordova:2016emh}, it can be shown that there can only be $D$-term deformations. Moreover, supersymmetry does not allow any 6-derivative interaction terms \cite{Cordova:2015fha}. Therefore, supersymmetry imposes that all local six derivative interactions of $\mathcal{L}_{\phi^4}$  must vanish implying the $a$-theorem\footnote{We follow the notation of reference \cite{Kundu:2019zsl} which  is different from the convention used in \cite{Cordova:2015fha}. In fact, one can use the equation $\Delta a_{\mathrm{tensor}} = \frac{2}{3} \hat{b}^2$ to relate our convention with that of \cite{Cordova:2015fha}. Also note that our $\hat{b}=b/f^2$ of \cite{Elvang:2012st}.}
\be\label{a:tensor}
a_{\rm UV}-a_{\rm IR}=\frac{2}{3}\hat{b}^2\ge 0\ .
\ee
The same conclusion can also be reached from the 5-pt effective interaction
\begin{align}
\mathcal{L}_{\phi^5}=&\frac{\hat{b}}{16f^8} \phi^3 \Box^2 \phi^2+\frac{1}{4f^{10}} \left(\Delta a-\frac{2}{3}\hat{b}^2\)\phi^3 \Box^3 \phi^2+\cdots \ . \label{int6}
\end{align}
In contrast, the $a$-theorem for tensor branch flows was derived in \cite{Cordova:2015fha} by studying 6-derivative 6-pt interactions. The same argument applies to the 6-pt effective interaction which implies $\Delta a \propto \hat{b}^2$ with a universal proportionality constant. The constant can be found to be $\frac{2}{3}$ (in our notation) by studying any special case.  This argument cannot provide the proportionality constant directly mainly because it is difficult to disentangle local and non-local parts of 6-derivative 6-pt interactions. On the other hand, our argument has the advantage that we can determine the constant of proportionality between $\Delta a$ and $\hat{b}^2$ directly from the effective action.

\subsection{Higgs Branch Flows}

Consider next the case of Higgs branch flows.  This is the branch of the moduli space parameterized at generic points by VEVs of hypermultiplets. One important feature of such flows is that they fully break the $SU(2)$ R-symmetry, and are often accompanied by Goldstone modes associated with the breaking of other global symmetries. An important remark is that the low energy effective action (\ref{eq:eff}) is general enough to capture all possible 6D Higgs branch flows since it does not even make explicit reference to supersymmetry, and just the notion of spontaneous breaking of an $SU(2)$ symmetry. The main point is that the terms of the dilaton effective action (\ref{eq:dilaton}) which involve just dilaton interactions are exactly the same as in the case of tensor branch flows, though in this case there can also be mixing terms with axions associated with global symmetry breaking. The interactions $\cL_{\rm mixed}[\phi,\xi]$ and $\cL_{\rm axion}[\xi]$ are given in Appendix \ref{app:DILAX}. Note that there should be other constraints on the remaining coupling constants from supersymmetry, however, we will not explore them here.

\subsection{Mixed Branch Flows}\label{ssec:MIXED}
This is the branch of the moduli space on which both tensor multiplets and hypermultiplets acquire VEVs. Hence, these flows also break the global $SU(2)_R$ symmetry. The effective field theory approach tells us something important about dilaton scattering amplitudes in 6D supersymmetric RG flows. In fact, the following analysis applies to any 6D RG flow with multiple symmetry breaking scales.

The main idea is to view a tensor branch flow and a Higgs branch flow as limiting cases of a mixed branch flow. Consider a 6D $\mathcal{N} =(1,0)$ SCFT in which conformal symmetry is broken by giving VEVs to a tensor multiplet scalar $t$ and a hypermultiplet scalars $H$ and $\widetilde{H}$:\footnote{In the above, $H$ and $\widetilde{H}$ denote complex scalars of (possibly different) hypermultiplets. There is a straightforward generalization to the case with multiple tensor multiplets and  hypermultiplets acquiring VEVs with different scales.}
\be\label{eq:vevs}
\langle t \rangle =f_t^2\ , \qquad \langle H\rangle =f_H^2 h\ , \qquad \langle \widetilde{H} \rangle = f_H^2 \widetilde{h}\ ,
\ee
where, the scales $f_t$ and $f_H$ have dimension of mass, whereas $h$ and $\widetilde{h}$ are dimensionless matrices (depending on the global symmetry breaking pattern) with $\cO(1)$ coefficients. This triggers an RG flow that preserves Poincar\'{e} supersymmetry
\begin{align}\label{eq:mixed}
{\rm SCFT_{ UV}}  \quad \Rightarrow \quad  &{\rm mixed\ branch} \nonumber\\*
&~~~~~~~\Downarrow \\*
&{\rm SCFT_{ IR}}  + S_{\rm eff}[\phi,\xi_a]\ ,\nonumber
 \end{align}
where, the effective action $S_{\rm eff}[\phi,\xi_a]$ is given by (\ref{eq:eff}). We can now imagine turning off the VEVs of all hypermultiplets $f_H=0$ and  hence, the flow (\ref{eq:mixed}) becomes a tensor branch flow. From the perspective of the low energy effective action $S_{\rm eff}[\phi,\xi_a]$, this corresponds to taking a  limit $\xi=0$. Furthermore, from the discussion of tensor branch flows we find that in this case
\be
\hat{b}= \hat{b}_{\rm tensor} \ , \qquad \Delta a= \Delta a|_{\rm tensor} =\frac{2}{3}\hat{b}_{\rm tensor}^2\ .
\ee
Similarly, one can consider the special case $f_t = 0$. Then the flow (\ref{eq:mixed}) becomes a Higgs branch flow
\be
\hat{b}= \hat{b}_{\rm Higgs} \ , \qquad \Delta a= \Delta a|_{\rm Higgs} \ .
\ee
We remark that sometimes it can happen that a tensor branch deformation can obstruct a Higgs branch deformation (the E-string theory
being a prominent example of this sort). This does not affect any of the statements presented here, since we have presupposed that we are dealing with a mixed branch flow in the first place.

\subsubsection{Effective Action}
Given an ${\rm SCFT_{ UV}} $ and VEVs (\ref{eq:vevs}), the dilaton and axions arise from the parameterization:
\be
t(x)= f_t^2 e^{-2\tau(x)}  \ , \qquad H(x) = f_H^2 e^{- 2\tau(x) } e^{-2 i  \beta_a(x) \sigma_a} h \ , \qquad \widetilde{H}(x) = f_H^2 e^{- 2\tau(x) } e^{-2 i  \beta_a(x) \sigma_a} \widetilde{h}\ ,
\ee
where $\tau(x)$ and $\beta(x)$ are the Goldstone modes of section \ref{sec:EFT} that nonlinearly realize the broken conformal snd $SU(2)_R$ symmetries. Again, here we have suppressed the contributions from other light scalars and axions. For example, in a flow which includes a Higgs branch deformation of a tensor branch deformed theory, we get a scalar for each tensor multiplet, and four real degrees of freedom (one quaternionic degree of freedom) from each hypermultiplet. The quantities $f_t$ and $f_H$ generate mass scales for states, and these massive modes must be integrated out to arrive at the corresponding effective action. The states parameterizing motion on the moduli space can be packaged in terms of massless tensor multiplets and hypermultiplets.

First, we focus on the tensor branch part of the flow and only integrate out modes that become massive because of the tensor multiplet VEV. This must coincide with the purely tensor branch flow and hence we can write the resulting effective action as
\be\label{tensor:action}
S_t= \int d^6x \(-2f_t^4e^{-4\tau}\(\p \tau\)^2+4\hat{b}_{\rm tensor} f_t^2 e^{-\tau}\Box^2 e^{-\tau} +3\Delta a|_{\rm tensor}  \tau \Box^3 \tau\)+\cdots\ ,
\ee
where, dots represent terms with 8 or more derivatives. Note that we are ignoring a possible cosmological constant term since it can be removed by adding a counterterm.

Next, we perform the same procedure but now for the hypermultiplets. We integrate out all  modes that become massive because of the hypermultiplet VEV(s). From the preceding section, we know that this procedure will generate the following effective action
\begin{align}\label{Higgs:action}
S_H=& \int d^6x \(-2f_H^4e^{-4\tau}\left(\(\p \tau\)^2+\frac{\gamma_0^2}{2} \Tr\(\p_\mu e^{i \beta} \p^\mu e^{-i \beta}\)\)+4\hat{b}_{\rm Higgs}  f_H^2 e^{-\tau}\Box^2 e^{-\tau} +3\Delta a|_{\rm Higgs}  \tau \Box^3 \tau\)\nonumber\\
&\qquad +\cdots\ ,
\end{align}
where dots represent terms involving axion interactions and terms with 8 or more derivatives. Finally, there can be additional terms with 4 or more derivatives that appear only in the mixed branch due to some intricate interactions between corresponding tensor and Higgs branches. Any such terms can be parametrized in the following way:
\be
S_{\rm mixed}= \int d^6x\(8\hat{b}_{\rm mixed}  f_H f_t e^{-\tau}\Box^2 e^{-\tau} +3\Delta a|_{\rm mixed}  \tau \Box^3 \tau\)+\cdots\ ,
\ee
where we are again ignoring higher-derivative terms and terms involving axions. Combining everything together and after a field redefinition
\be\label{eq:rescaled}
e^{- 2\tau } =1-\frac{\phi}{f^2}+\cdots\ ,
\ee
where dots represent terms involving axions, we must recover the effective action (\ref{eq:eff}). In particular, at energy scales $E\ll f_t,f_H$ we obtain the dilaton effective action (\ref{eq:dilaton}) once we identify
\begin{align}
&f^4= f_t^4+f_H^4\ , \\
&\hat{b}=\frac{1}{f^2}\(f_t^2 \hat{b}_{\rm tensor} +2f_t f_H \hat{b}_{\rm mixed} +f_H^2\hat{b}_{\rm Higgs} \)\ ,\label{b:mixed}\\
&\Delta a= \Delta a|_{\rm tensor}+\Delta a|_{\rm mixed}+\Delta a|_{\rm Higgs} \label{a:mixed}\ .
\end{align}
Note that $\hat{b}_{\rm tensor}$, $\hat{b}_{\rm Higgs}$, $\Delta a|_{\rm tensor}$, and $\Delta a|_{\rm Higgs}$ do not depend on the scales $f_t$ and $f_H$. On the other hand, both $\hat{b}_{\rm mixed}$ and $\Delta a|_{\rm mixed}$ can depend on the ratio $f_t/f_H$. Moreover, we make the mild assumption that $\hat{b}_{\mathrm{mixed}}$ is non-singular and hence in both limits $f_t\gg f_H\gg E$ and $f_H\gg f_t \gg E$
\be
\frac{f_t f_H}{f^2}\hat{b}_{\rm mixed}\rightarrow 0\ .
\ee

\begin{figure}
\begin{center}
\includegraphics[scale = 0.55, trim = {0cm 3.0cm 0cm 5.0cm}]{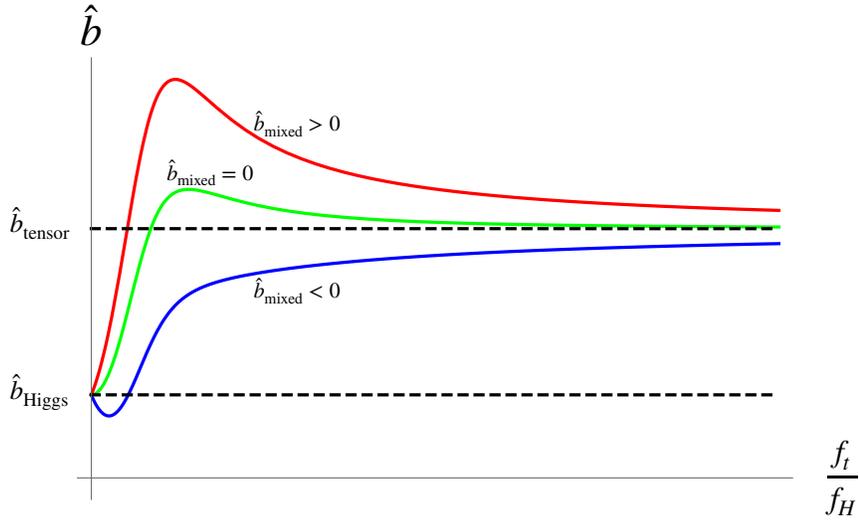}
\end{center}
\caption{\small A schematic plot of how the parameter $\hat{b}$ depends on $f_t/f_H$ in a 6D mixed branch flow. The full $\hat{b}$ interpolates between $\hat{b}_{\rm Higgs}$ and $\hat{b}_{\rm tensor}$ as we increase $f_t/f_H$. Three lines in the plot represent three special cases: $\hat{b}_{\rm mixed}<0$ (blue), $\hat{b}_{\rm mixed}=0$ (green), and $\hat{b}_{\rm mixed}>0$ (red). Note that $\hat{b}$ has a non-trivial dependence on $f_t/f_H$ even when a mixed branch flow factorizes completely ($\hat{b}_{\rm mixed}=0$).}%
\label{fig:plotb}%
\end{figure}

\subsubsection{Asymptotics}
The main point to take away from the above discussion is that both $\hat{b}$ and $\Delta a$ can depend on two scales $f_t$ and $f_H$ in a very specific way, as shown in figure \ref{fig:plotb}. In particular, $\hat{b}$  interpolates between
\be
 f_t \gg f_H: \qquad \hat{b}=\hat{b}_{\rm tensor} +2\(\frac{f_H}{f_t}\)\hat{b}_{\rm mixed}(f_H=0)+\cdots\ ,
\ee
and
\be
f_H \gg f_t: \qquad \hat{b}=\hat{b}_{\rm Higgs} +2\(\frac{f_t}{f_H}\)\hat{b}_{\rm mixed}(f_t=0)+\cdots\  .
\ee
This implies that if $\hat{b}$ is known as a function of $f_t/f_H$,
we can recover $\hat{b}_{\rm tensor}$ and $\hat{b}_{\rm Higgs}$ from their asymptotic values.

\subsubsection{Mixed Contributions and Factorization}
Let us end this section by providing a physical explanation of the relations (\ref{b:mixed}) and (\ref{a:mixed}). Consider the dilaton 4-point amplitude $\cA(\phi\phi\phi\phi)$. First, we give a VEV to only the tensor multiplet scalar $t$. The dilaton 4-point amplitude $\cA(\phi\phi\phi\phi)|_{\rm tensor}$ is obtained by adding all Feynman diagrams with 4 external $\phi$ with massive states (due to the tensor multiplet VEV) running inside. This amplitude is given by
\be\label{amplitude:tensor}
\cA(\phi\phi\phi\phi)|_{\rm tensor}= \frac{\hat{b}_{\rm tensor}}{2f_t^6}\(s^2+t^2+u^2\)+\frac{9}{2f_t^8}\(\Delta a|_{\rm tensor}-\frac{2}{3}\hat{b}_{\rm tensor}^2\)stu + \cdots\ ,
\ee
where the $stu$ term vanishes because of the condition (\ref{a:tensor}). Similarly, for the Higgs branch, the dilaton 4-point amplitude $\cA(\phi\phi\phi\phi)|_{\rm Higgs}$ is obtained by adding all Feynman diagrams with 4 external $\phi$ with massive states (due to the VEVs which break R-symmetry) running inside
\be\label{amplitude:higgs}
\cA(\phi\phi\phi\phi)|_{\rm Higgs}= \frac{\hat{b}_{\rm Higgs}}{2f_H^6}\(s^2+t^2+u^2\)+\frac{9}{2f_H^8}\(\Delta a|_{\rm Higgs}-\frac{2}{3}\hat{b}_{\rm Higgs}^2\)stu+\cdots\ .
\ee
When we combine both VEVs to obtain a mixed branch flow, the full dilaton 4-point amplitude $\cA(\phi\phi\phi\phi)$ receives contributions from both of the above processes. However, in the mixed branch flow both amplitudes (\ref{amplitude:tensor}) and (\ref{amplitude:higgs}) must be rescaled. This is because of the field redefinition (\ref{eq:rescaled}) which involves the scale $f$. Hence, the full amplitude is given by
\be
\cA(\phi\phi\phi\phi) = \(\frac{f_t}{f}\)^8 \cA(\phi\phi\phi\phi)|_{\rm tensor}+ \(\frac{f_H}{f}\)^8 \cA(\phi\phi\phi\phi)|_{\rm Higgs}+\cA(\phi\phi\phi\phi)|_{\rm mixed}\ ,
\ee
where $\cA(\phi\phi\phi\phi)|_{\rm mixed}$ represents Feynman diagrams with 4 external $\phi$ with massive states from both tensor and Higgs branches running inside the loops. These diagrams also include massive states running inside the loops with masses that depend on both mass scales $f_t$ and $f_H$. The above amplitude is completely consistent with the relations (\ref{b:mixed}) and (\ref{a:mixed}). Moreover, the above amplitude immediately implies that both $\hat{b}_{\rm mixed} $ and $\Delta a|_{\rm mixed}$ arise only from Feynman diagrams with massive states from both tensor and Higgs branches running inside. The same conclusion can also be reached by studying 5-pt, and 6-pt dilaton amplitudes.

A mixed branch flow factorizes completely when $\cA(\phi\phi\phi\phi)|_{\rm mixed}=0$. In that case, the mixed branch flow is a combination of a purely tensor branch flow and a purely Higgs branch flow. Another important special case is the partial factorization $\Delta a|_{\rm mixed}=0$. In other words, the IR fixed point does not depend on the scales $f_t$ or $f_H$ as long as both of them are non-zero. In this case, we can take different scaling limits of $f_t/f_H$ of the RG flow without affecting $\Delta a$.


\section{6D\ SCFT Mixed Branch Flow Examples}\label{sec:example}
In the previous section we showed that our dilaton-axion effective action
applies even when there are multiple sources of conformal symmetry breaking.
In this section we illustrate some of these general features with an explicit
example based on 6D SCFTs. Recall that in 6D\ SCFTs, there are generically two
sources of conformal symmetry breaking. One comes from motion on the tensor
branch of an SCFT, and the other comes from motion on the Higgs branch. In
general, a mixed branch flow can be decomposed into an alternating sequence of
Higgs branch and tensor branch deformations. We would in particular like to
extract the contribution to 4-point dilaton scattering from the term:%
\begin{equation}
L\supset\frac{\hat{b}}{16f^{6}}\phi^{2}\Box^{2}\phi^{2}%
\end{equation}
due to motion on the tensor branch, as well as the Higgs branch.

In general, this is a challenging question to answer because the strongly
coupled nature of all 6D\ SCFTs makes it difficult to provide a microscopic
analysis of how such terms arise. That being said, we can still make some
general comments. First of all, when we move onto the tensor branch, we have
effective strings in the low energy theory, with a tension which scales as:%
\begin{equation}
T_{\text{eff}}\sim f_t^{2},
\end{equation}
with $f_t$ a \textquotedblleft decay constant\textquotedblright\ energy
scale associated with the VEV of the scalar in a tensor multiplet (\ref{eq:vevs}). The origin
of the moduli space is where these effective strings would appear to have
vanishing tension. Rather than signalling the appearance of a non-local
theory, this instead points to the appearance of a strongly coupled conformal
fixed point \cite{Seiberg:1996qx}. In fact, all known $D>4$ CFTs have the
property that effective strings have vanishing tension at a fixed point. This
is rather different from what happens in the better studied case of $D\leq4$ CFTs, where
the approach to a conformal fixed point involves vanishing masses for effective particles.

Indeed, the general method for constructing 6D\ SCFTs involves starting on the tensor
branch of a candidate 6D\ SCFT, and then attempting to pass to the origin of
tensor branch moduli space. In such cases, it is often possible to perform a
further deformation onto a Higgs branch, though sometimes there can be
obstructions. As an example where there is an obstruction, consider the
E-string theory obtained from an M5-brane probing a heterotic $E_{8}$
9-brane. In this case, the Higgs branch of the 6D\ SCFT corresponds to
dissolving the M5-brane as flux (an instanton)\ in the nine-brane, while the
tensor branch deformation corresponds to pulling the M5-brane off the wall,
completely destroying the existence of a Higgs branch.

But there are also cases where no obstruction is generated
\cite{Heckman:2016ssk}. For example, consider
$\mathfrak{su}(N)$ gauge theory coupled to $2N$ hypermultiplets in the
fundamental representation. In this case, anomaly cancellation requires
coupling the non-abelian vector multiplet to a single tensor multiplet of
charge $-2$. In F-theory terms this 6D\ SCFT is realized by a curve of
self-intersection $-2$ with an $I_{N}$ fiber over the $-2$ curve which
collides with $2N$ fibers with an $I_{1}$ singularity. These flavor symmetry
fibers can be tuned to make manifest an $\mathfrak{su}(2N)$ flavor symmetry
algebra (in which case we have a single $I_{2N}$ fiber which collides with the
$I_{N}$ fiber).

For our present purposes the key point is that all of
the Higgs branch deformations of the 6D\ SCFT are captured by deformations
which are directly visible in terms of explicit VEVs for weakly coupled
hypermultiplets. As an illustrative example, the deformation from a theory
with $\mathfrak{su}(N)\ $gauge symmetry to $\mathfrak{su}(N-1)$ gauge symmetry
involves giving a VEV to precisely two flavors from the system, namely we wind
up with $2N-2$ flavors from such a Higgs branch deformation. This is in accord
with the fact that to satisfy the triplet of D-term constraints in the gauge
theory, a pair of hypermultiplets actually needs to get a VEV
\cite{Danielsson:1996uv, Morrison:2012np}. More generally, all these Higgs
branch deformations correspond to brane recombination moves
\cite{Hassler:2019eso}. Another interesting aspect of this example is that
upon compactification of the tensor branch theory on a $T^{2}$, we obtain a 4D
$\mathcal{N}=2$ SCFT with the same gauge group and matter content. In this
model, the reduction of the tensor multiplet decouples, which fits
with the fact that effective strings play little role in the approach to the
4D\ conformal fixed point.

We would now like to better understand the possible contributions on the Higgs
branch to 4-dilaton, 4-derivative interaction terms. As a warmup
exercise, let us first consider the case of the 4D\ $\mathcal{N}=2$ SCFT
obtained from dimensional reduction of the tensor branch of the 6D\ SCFT. In
this setting, we have a marginal gauge coupling constant which we can use to
tune the theory to weak coupling. Carrying out the analysis there, we are free
to parameterize possible contributions to $SU(2)\times U(1)$ R-symmetry
breaking of the 4D\ theory in terms of a sigma model field $\Sigma$ which we
schematically write as:%
\begin{equation}
\Sigma\sim f\exp\left(  -\tau+\text{axions}\right)\ , \qquad \exp(-\tau)=1- \frac{\phi}{f}+\cdots\ ,
\end{equation}
where in the case at hand, the $\Sigma$'s are related to the hypermultiplets.
We wish to track possible contributions to 4-dilaton, 4-derivative terms, and so the general structure in the 4D Lagrangian for
the $\Sigma$ fields will involve terms of the schematic form:\footnote{Of course, $\hat{b}$ is actually $\Delta a$ in 4D. }%
\begin{equation}
L_{4D}\supset\frac{1}{\left(  \Lambda_{4D}\right)  ^{2n}}%
g^{2m}\partial^{4} \Sigma^{2n}\Rightarrow \frac{\Delta a}{2f^{4}}\phi^{2}\Box^{2}\phi^{2},
\end{equation}
where $m$ and $n$ are non-negative integers and $g$ is the 4D gauge coupling. Here, we have schematically indicated the number of powers of
derivatives, and have also left open the different possible ways that the $\Sigma$ fields may actually multiply together, suppressing their hermitian conjugates. Indeed, the only condition we are actually imposing is that the number of $\Sigma$ fields is even. This requirement follows in the effective field theory from the symmetry $\Sigma \mapsto - \Sigma$.\footnote{At the origin of the Higgs branch we have a $\mathbb{Z}_2$ symmetry, and at more generic points of the Higgs branch, we still parameterize fluctuations around a background VEV in terms of hypermultiplets, so there is still an unbroken $\mathbb{Z}_2$ symmetry.} It can also be seen directly by observing that in any non-zero dilaton scattering amplitude, an even number of hypermultiplets must appear. The power of $g$ is also an even number, as can be verified by working with a basis of fields where all $g^2$ dependence is in the vector boson propagator.

\begin{figure}[t]
\begin{center}
\includegraphics[scale = 0.5, trim = {0cm 2.0cm 0cm 2.0cm}]{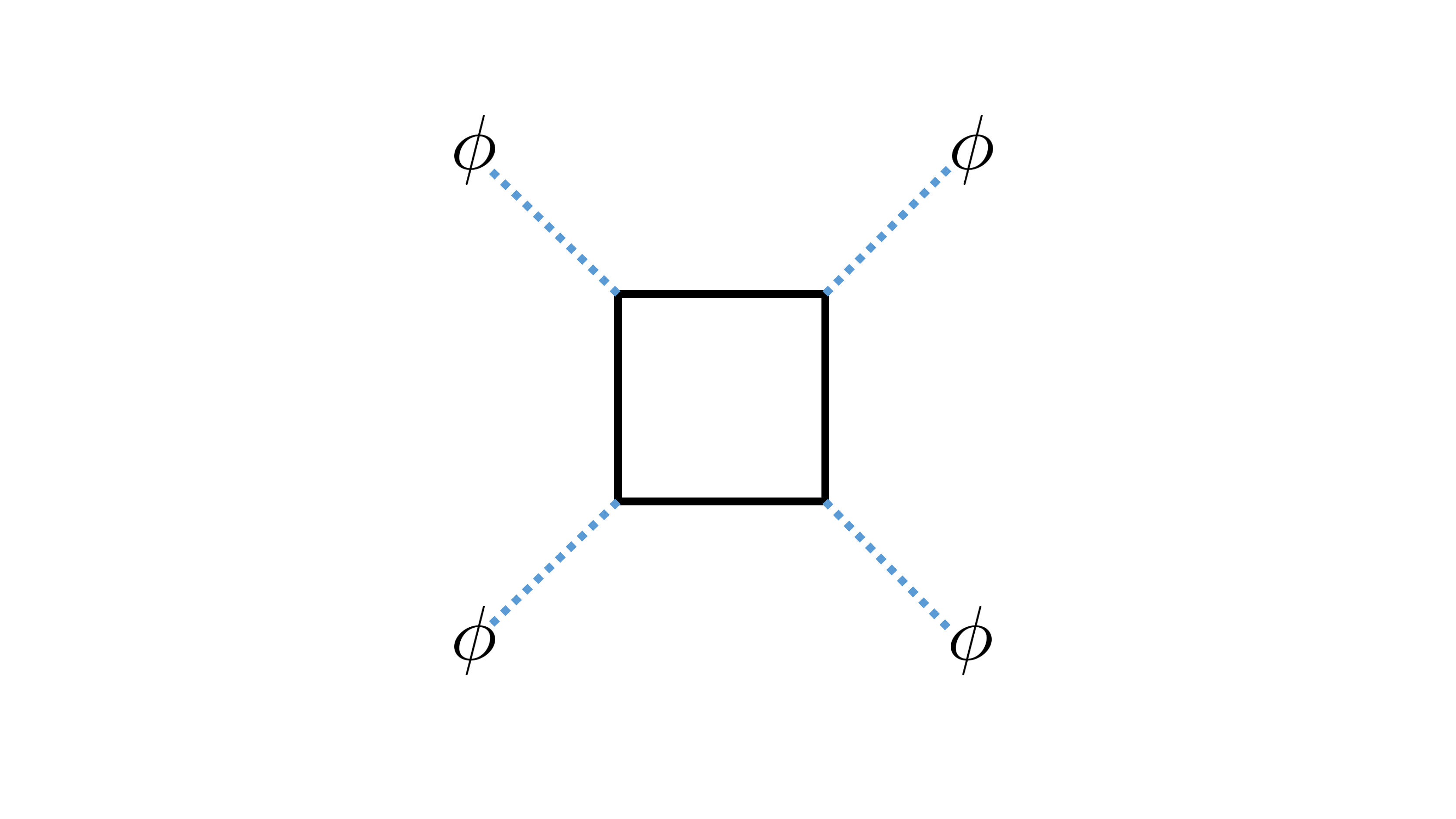}
\end{center}
\caption{\small Example of a 1-loop diagram which contributes to 4-dilaton
4-derivative interaction terms in the dilaton effective action obtained from a
Higgs branch flow. Here, we have indicated external scalars as associated with
the hypermultiplets, with internal massive gauginos running in the loop. In
4D, this diagram scales as $\mathcal{M}_{2\rightarrow2}\sim g^{4}p^{4}/\left(
\Lambda_{4D}\right)  ^{4}=p^{4}/f_{H}^{4}$, while in 6D\ we instead find
$\mathcal{M}_{2\rightarrow2}\sim g^{4}p^{4}/\left(  \Lambda_{6D}\right)
^{2}=p^{4}/f_t^{2}f_{H}^{4}$, with $p$ the characteristic momentum
of the external states.}%
\label{fig:4DilatonScattering}%
\end{figure}

From this starting point we get interaction terms for the dilaton and the axions by expanding
the exponential $\Sigma$ in powers of $\phi/f$, where $\phi$ is the physical dilaton which is related to $\tau$ in the usual way \cite{Kundu:2020bdn}. The mass scale
$\Lambda_{4D}$ is not arbitrary, and is set by the masses of vector bosons
obtained from actually Higgsing the $\mathfrak{su}(N)$ gauge symmetry. In
particular, we have the further relation:%
\begin{equation}
\left(  \Lambda_{4D}\right)  ^{2}\sim g^{2}f^{2}.
\end{equation}
Plugging in, the form of our coupling in the 4D\ Lagrangian is:%
\begin{equation}
L_{4D}\supset\frac{1}{f^{2n}}g^{2m-2n}\partial^{4} \Sigma^{2n} \Rightarrow \frac{\Delta a}{2f^{4}}\phi^{2}\Box^{2}\phi^{2}\ .
\end{equation}
So, we see that we can generate contributions with suppression scale set
purely by $f$ (the Higgs branch scale) just by setting $m=n$. An explicit
example of a 4D\ loop correction which generates this sort of structure is
shown in figure \ref{fig:4DilatonScattering}.

Let us now turn to the analog of these expressions for our 6D\ theory. In this
case, a Higgs branch deformation of the tensor branch deformed theory will
necessarily need to make reference to at least two mass scales, namely the
tensor branch deformation scale $f_t$ as well as the Higgs branch
deformation scale $f_{H}$, as discussed in the previous section (see equation (\ref{eq:vevs})).
We expect a gauge theory description to provide an adequate approximation provided we work at energy scales with $s<f_t^{2}$.
Self-consistency of the Higgs mechanism in this regime also requires
$f_{H}<f_t$. Finally, observe that the gauge coupling is no longer
dimensionless, but instead scales as $g^{2}\sim1/f_t^{2}$. As a general comment, we note that even though we will perform
our analysis in the regime $f_H < f_t$, there is no hint in the F-theory geometry of a phase transition taking place as we pass to the regime
$f_t < f_H$. This strongly indicates that results obtained in one regime of validity
do not suddenly break down in other regions of the moduli space.

Before we proceed some comments are in order. In the low energy effective action there is no clear way to distinguish between $\hat{b}_{\rm Higgs}$ and $\hat{b}_{\rm mixed}$ (or $\Delta a|_{\rm Higgs}$ and $\Delta a|_{\rm mixed}$). However, we can define an effective $\hat{b}_{\rm Higgs}$ in the following way.\footnote{One can define an effective $\Delta a|_{\rm Higgs}$ as well, however, we are mainly interested in  $\hat{b}_{\rm Higgs}$ since it is the leading interaction in the effective theory.}  First, consider the purely tensor branch flow $f_H=0$. Now we perform a Higgs branch deformation of the tensor branch deformed theory: $f_H\ll f_t$. One can then ask the question whether the resulting RG flow, at the leading order, can be though of as a combination of a purely tensor branch flow and a purely Higgs branch flow (or equivalently a factorized mixed branch flow, as defined in the previous section). This necessarily requires that $\hat{b}$ does not have a linear $f_t$ contribution when $f_H\neq 0$, as seen from equation (\ref{b:mixed}).\footnote{There is an additional requirement that the correction to $\Delta a$ from the Higgs branch deformation is independent of $f_t$ at the leading order in $\frac{f_H}{f_t}$. } If this condition is satisfied, we can define an effective $\hat{b}_{\rm Higgs}$:
\be\label{effective:b}
\hat{b}_{\rm Higgs}= \frac{1}{2}\(\frac{\p^2\ \hat{b} f^2}{\p f_H^2}\)_{\frac{f_H}{f_t}\rightarrow 0}\ .
\ee
So, in this case the mixed branch flow at the leading order can be effectively described by the low energy effective action
\be
S_{\rm eff}[\phi,\xi]=S_t+S_H\ ,
\ee
where $S_t$ is the effective action (\ref{tensor:action}) for the tensor branch
flow and $S_H$ is the action (\ref{Higgs:action}) associated with a purely Higgs
branch flow with $\hat{b}_{\rm Higgs}$ given by (\ref{effective:b}).

Let us now track the 6D analogs of the contributions to the dilaton effective
action identified in 4D and compute the effective $\hat{b}_{\rm Higgs}$. In 6D, the sigma model field $\Sigma$ of R-symmetry
breaking can be schematically written as:%
\begin{equation}
\Sigma\sim f_{H}^{2}\exp\left(  -2\tau+\text{axions}\right)  \ , \qquad \exp(-2\tau)=1- \frac{\phi}{f^{2}}+\cdots\ .
\end{equation}
Here, we must distinguish $f_{H}$, the scale associated with R-symmetry
breaking, with the more general scale $f$ which appears in our discussion of
mixed-scale conformal symmetry breaking. Consider, then, terms that can generate 4-dilaton 4-derivative interactions of the form:
\begin{equation}\label{eq:L6D}
L_{6D}\supset\frac{1}{(\Lambda_{6D})^{4n-2m-2}}g^{2m}\partial
^{4} \Sigma^{2n} \Rightarrow \frac{\hat{b}}{16f^{6}}\phi^{2}\Box^{2}\phi^{2}.
\end{equation}
Again, the nature of R-symmetry breaking restricts us to an even number of powers of $\Sigma$ fields, but other than that, we
have written down the general form of the possible higher-dimension operator.
In the above, $\Lambda_{6D}$ is a scale associated with the massive vector bosons:
\begin{equation}
\left(  \Lambda_{6D}\right)  ^{2}\sim g^{2}f_{H}^{4}=f_{H}^{4}/f_t^{2}.
\end{equation}
Now  by expanding  (\ref{eq:L6D}) we see that
\be
\hat{b}\sim \frac{f_H^{4n}}{f^2} \frac{g^{2m}}{(\Lambda_{6D})^{4n-2m-2}}\ .
\ee
We are specifically interested in a purely Higgs branch contribution to $\hat{b}$, as defined in (\ref{effective:b}).
Equation (\ref{b:mixed}) implies that $\hat{b}f^2$ for such a contribution must be independent of $f_t$, up to terms that are suppressed for $f_H/f_t\ll 1$.
So, we are interested in generating higher-dimension
operators that lead to $f_t$ independent contributions to $\hat{b}f^2$,
which in turn requires a specific power of $m$.
In particular, we need:%
\begin{equation}
2m-(4n-2m-2)=0,
\end{equation}
or:%
\begin{equation}
2n-2m=1,
\end{equation}
which cannot be arranged if $m$ and $n$ are integers. So, the first
distinction from the seemingly related 4D case is that the same sorts of
diagrams used in 4D will \textit{not} produce a purely $f_{H}$ dependent contribution to $\hat{b}f^2$  in 6D.
Rather, they will involve a combination of
scales including both $f_{H}$ and $f_t$.

We next consider a slightly broader class of higher-dimension operators and show again
that $\hat{b}f^2$ can only have terms that are suppressed by positive powers of $1/f_t$.
In a mixed branch flow the dilaton will be a linear
combination of contributions coming from the scalars of both the
hypermultiplet as well as the tensor multiplet. One might ask whether we
can generate a contribution of the desired
form using this larger set of fields. To this end, write:
\begin{equation}\label{tensorexpand}
t\sim f_t^{2}\exp\left( -2\tau\right)  \ , \qquad \exp(-2\tau)=1- \frac{\phi}{f^{2}}+\cdots
\end{equation}
so we could in principle track more general combinations involving powers of
$t$ and $\Sigma$. The corresponding terms of interest in the Lagrangian are:
\begin{equation}\label{eq:L6Dagain}
L_{6D}\supset\frac{1}{(\Lambda_{6D})^{4n + 2l -2m-2}}g^{2m}\partial
^{4} \Sigma^{2n} t^{l} \Rightarrow \frac{\hat{b}}{16f^{6}}\phi^{2}\Box^{2}\phi^{2},
\end{equation}
with $m$, $n$ and $l$ non-negative integers. The appearance of suppression scales can be argued for as follows. First, we
move onto the tensor branch, and integrate out all massive states (including those coming from excitations of the effective strings).
This puts us on the tensor branch gauge theory, with light degrees of freedom given by the vector multiplet, tensor multiplet and hypermultiplets. In this theory, the VEVs for hypermultiplets generate masses as specified by $\Lambda_{6D}$. In particular, we need not concern ourselves with subleading masses generated by hypermultiplet / effective string couplings, since we have already integrated out the degrees of freedom associated with string excitations.

We would again like to figure out possible contributions to $\hat{b} f^2$ which can come from just the scale
$f_H$. In this case, we must be mindful of the fact that in extracting the interaction terms for the dilaton,
we can also expand the tensor multiplet scalar, as in equation (\ref{tensorexpand}), yielding
\be\label{b:perturbative}
\hat{b}\sim \frac{f_H^{4n}}{f^2} \frac{f_{t}^{2l} g^{2m}}{(\Lambda_{6D})^{4n+2l-2m-2}}\ .
\ee
 Let us now see whether
adding in insertions of the scalars $t$ can help us generate a term in $\hat{b}f^2$ which is independent of $f_t$.
The condition we now have to satisfy is:
\begin{equation}\label{eqn:condition}
2l = 2m - 2n + 1.
\end{equation}
Again, this equation cannot be satisfied for integer $l,m,n$, so we arrive at a contradiction.
From this, we conclude that at least at the level of explicit Feynman diagrams,
it is rather challenging to identify an explicit contribution to 4-dilaton, 4-derivative interactions with Wilson coefficient $\sim f_H^2/f^8$.

We can also try to place some constraints on how such strong coupling effects might enter.
Note that we are in the regime $s < f^2_H < f^2_t$ where a gauge theory description
should be valid. Adding more loops simply corresponds to increasing the power of $g^2$ in a contribution to the
dilaton effective action. In fact, we can see that the number of powers of $g^2$ is bounded below by the number of external legs
coming from $\Sigma$ and $t$. To derive such a bound, it is convenient to
canonically normalize the gauge fields so that no factors of $g^2$ appear
in their propagators. Observe that $t$ mixes with the gauge fields via the interaction term $t \mathrm{Tr} F^2$
of the tensor branch Lagrangian. Canonically
rescaling the gauge fields amounts to the operation:
\begin{equation}
t \mathrm{Tr} F^2 \mapsto \frac{t}{f_t^2} \mathrm{Tr} F^2.
\end{equation}
Consider next the insertions of the $\Sigma$ fields, or equivalently, the presence of the
hypermultiplet scalars $H$ (and their complex partners $\widetilde{H}$), expanded around a background VEV.
We have two sorts of interaction terms, those which go as $H^\dag H A^2$ (quartic vertex of order $g^2$), and those which go as $H^{\dag} \partial H A - h.c.$. (cubic vertices of order $g$). Here, $A$ denotes a vector boson. There are also supersymmetric analogs of these interaction terms, involving the fermionic superpartners as well as the D-term potential(s) of the vector multiplet. Our claim is that each such insertion of an external $H$ field is accompanied by a power of $g$. The main point is that at least for dilaton scattering, the relevant contribution from a term such as $H^{\dag} \partial H A$ will necessarily involve a coupling between a dilaton, an eaten (i.e. massive) Goldstone boson, and the (massive) gauge boson. The same considerations apply for the Yukawa interaction terms required by supersymmetry, where all fermions are necessarily internal anyway.\footnote{One might ask whether the quartic terms associated with the D-term potential could generate interaction terms with lower powers of $g^2$. The supersymmetric Higgs mechanism excludes such a possibility.} From this, we conclude that to each such interaction term involving a $\Sigma$ insertion, we can associate at least one power of $g$. Putting these two considerations together, we conclude that in the operator of line (\ref{eq:L6Dagain}) where we have $l$ insertions of $t$, and $2n$ insertions of $\Sigma$, we have the inequality:
\begin{equation}\label{eqn:boundo}
m \geq n + l.
\end{equation}
In particular, this means that as we move to higher loop order, we simply generate further powers of $f_t$ in the denominator.\footnote{Note that the inequality of line (\ref{eqn:boundo}) provides an alternative way to exclude higher-dimension operators which makes no reference to the integrality of various exponents. For example, returning to equation (\ref{eqn:condition}), we get: $2l = 2m - 2n + 1 \geq 2(n+l) - 2n + 1 = 2l + 1$, a contradiction.}

Putting all of this together, we observe from equation (\ref{b:perturbative}) that the general structure of the contributions to $\hat{b}$ take the form of a power series
in the ratio $r = f^2_H / f^2_t$:\footnote{In fact, even more is true, only the coefficients 
$\hat{b}_{m}$ with $m = 2k + 1$ for $k \in \mathbb{Z}_{\geq 0}$ are non-zero.}
\begin{equation}\label{result:b}
f^2 \hat{b}(f_H, f_t) =f_t^2 \hat{b}_{\rm tensor}+f_H^2 \underset{m > 0}{\sum} \hat{b}_{m} r^{m},
\end{equation}
namely the corrections to $f^2 \hat{b}$ from the Higgs branch always have some suppression by $f_t$. In particular, note that there is no contribution which is linear in $f_t$. Hence, we can describe this mixed branch flow, at the leading order, as a combination of a purely tensor branch flow and a purely Higgs branch flow. Furthermore, the effective $\hat{b}_{\rm Higgs}$, as defined in (\ref{effective:b}), is given by:
\be\label{b:zero}
\hat{b}_{\rm Higgs}=0\ .
\ee
As we increase $f_H/f_t$, the second term on the righthand side of (\ref{result:b}) starts contributing. However, since the second term depends on $r = f^2_H / f^2_t$, it should be interpreted as a contribution to $\hat{b}_{\rm mixed}$, as defined in (\ref{b:mixed}).

Summarizing our discussion up to this point, we have seen that at least in the
regime where $f_{H}<f_t$, the contributions to $\hat{b}$ are rather
different in 6D\ compared with their 4D counterpart. In fact, the above
analysis would seem to imply that at leading order in $f_H/f_t$,
there are actually no contributions to $\hat{b}_{\mathrm{Higgs}}$!

Suppose we now attempt to extrapolate to the other regime where $f^2_H / f^2_t$ is large, namely, the limit where we approach a pure Higgs branch flow. From our expansion, we can see that individual terms would appear to diverge (since we are taking small $1/r$ now). On the other hand, another analytic expansion in $1/r$ must be available simply because we have assumed the existence of an SCFT in the first place! This means that there is a re-summation of these terms available which should in principle allow us to analytically continue to the other regime. In fact, the absence of any hint of a phase transition in the F-theory geometry when $0 < r < \infty $ again strongly suggests that such a re-summation must exist. Moreover, as we explained earlier, this re-summation of equation (\ref{result:b}) must be interpreted as contributing to $\hat{b}_{\rm mixed}$. With this in mind, there still does not appear to be a natural candidate for generating an effective $\hat{b}_{\rm Higgs}$  contribution to 4-pt 4-derivative dilaton scattering, even in the limit $f_t \rightarrow 0$.

Caution is of course warranted for the following reason. In the regime $f_H > f_t$,
any number of things could happen, since the scale of R-symmetry breaking is now bigger than that
of the mass scale set by $f_t$. A priori, it could happen that all of the
above analysis is invalidated, though this would require the onset of a phase
transition as a function of $f_{H}/f_t$ at some finite non-zero value, a feature which
is wholly absent from any known 6D\ SCFT construction. In particular, in all F-theory constructions, the
complex structure and K\"{a}hler deformations can (when not obstructed) be
scaled independently of one another. Provided we remain in the regime $s\ll
f_t^{2}\ll f_{H}^{2}$, it seems implausible that some unknown
\textquotedblleft strong coupling effect\textquotedblright\ completely
destroys the scaling estimates just presented.

We expect similar considerations to apply, even in situations where
no pure gauge theory description of a Higgs branch deformation is available,
as well as cases in which part of the Higgs branch of the 6D SCFT is obstructed by moving onto the tensor branch.
As an example, consider again the E-string theory. This theory can
be viewed as part of a sequence of theories indexed by $n \in \mathbb{Z}_{\geq 0}$,
with an $\mathfrak{sp}(n)$ gauge algebra coupled to a tensor multiplet with charge $-1$.
In this case, the F-theory construction makes manifest an $\mathfrak{so}(2n+8)\times
\mathfrak{so}(2n+8)$ flavor symmetry (which enhances further at the origin of
the tensor branch). The resulting \textquotedblleft analytic
continuation\textquotedblright\ in the value of $n$ to $n=0$ has been observed
to be compatible with the structure of anomalies in these theories
\cite{Heckman:2016ssk, Heckman:2018pqx}. In all these cases,
we again expect that the mass scale for the corresponding ``vector bosons''
is of the form $\left(  \Lambda_{6D}\right)  ^{2}\sim g^{2}f_{H}^{4}=f_{H}^{4}/f_t^{2}$.

These considerations are corroborated in the F-theory
geometry as follows. A Higgs branch deformation is associated with
a complex structure deformation of the elliptically fibered Calabi-Yau threefold, and as such, is
correlated with a compact 3-cycle. Each such 3-cycle can be understood as a two-cycle in the F-theory
model fibered over a 1-chain. Geometrically, the volumes are related as $\mathrm{Vol(3-cycle)} \sim \mathrm{Vol(1-chain)} \times \mathrm{Vol(2-cycle)}$, which is just the relation $f_{H}^2 \sim g^{-1} \Lambda_{6D}$, namely, $f_{H}^2 \sim \mathrm{Vol(3-cycle)}$,
$g^{-1} \sim \mathrm{Vol(1-chain)}$ and $\mathrm{Vol(2-cycle)} \sim \Lambda_{6D}$.

Given our difficulties with generating a contribution to $\hat{b}_{\mathrm{Higgs}}$ in 6D SCFTs,
one might now ask whether similar considerations apply in the behavior of
$\Delta a$. For 4D Higgs branch flows, there is no issue, because
$\hat{b}$ and $\Delta a$ are proportional to one another, both being dictated by
the $\phi^{2}\Box^{2}\phi^{2}$ interaction term of the dilaton
effective action. For 6D\ Higgs branch flows, however, the leading order
contribution to $\Delta a$ comes from 6-derivative terms such as $\phi^2 \Box^3 \phi^2$.
Running through the same line of reasoning just presented, we now seek out terms of the form:
\begin{equation}
L_{6D}\supset\frac{1}{(\Lambda_{6D})^{4n-2m}}g^{2m}\partial
^{6} \Sigma^{2n} \Rightarrow \frac{3\Delta a}{16f^8} \phi^2 \Box^3 \phi^2.
\end{equation}
In this case, the condition to eliminate all $f_t$ dependence from the
suppression scale amounts to the condition:%
\begin{equation}
4n-4m=0,
\end{equation}
which can be satisfied for integer values of $m$ and $n$. Note that this makes the result (\ref{b:zero}) rather important. This implies that this class of 6D mixed branch flows, at the leading order, can be described as a combination of a purely tensor branch flow and a purely Higgs branch flow, where the Higgs branch part is actually non-trivial since $\Delta a|_{\rm Higgs}$ may not be suppressed by $f_H/f_t$.

\section{Conjecture on $\hat{b}_{\text{Higgs}} = 0$ \label{sec:CONJ}}

The above considerations when combined with our discussion from the previous section, would seem to indicate
that $\hat{b}_{\text{Higgs}}=0$, suggesting a substantial error in our seemingly naive
analysis, since it stands in contrast to a huge body of literature. That being said,
our analysis of the dilaton effective action suggests that generating a $\phi^2 \Box^2 \phi^2$
interaction term is actually rather non-trivial. In this section we present a conjecture that
$\hat{b}_{\text{Higgs}}$ is, in fact zero, explaining some of the evidence in favor of such a surprising sounding statement.

Indeed, there are a few important caveats to keep in mind in asserting that $\hat{b}%
_{\text{Higgs}}=0$ cannot occur in an interacting theory. In the analysis of
\cite{Adams:2006sv}, the condition $\hat{b}>0$ for an interacting theory
relies on the assumption that at large $s$, there are no singularities present
in 4-point dilaton scattering in the large $s$ regime. This is a
natural condition to impose, and in $D\leq4$ theories it is very well
supported. Conformal bootstrap methods also establish $\hat{b}\geq0$ \cite{Kundu:2019zsl}. To
date, there has been far less attention paid to the case of $D>4$ CFTs, where
in all known examples, effective strings are crucial for the existence of a
fixed point in the first place.

To begin, observe that the tree-level $2\rightarrow 2$ amplitude of the dilaton for any 6D RG flow has the form:
\begin{align}
\cA_4(s,t)=\frac{\hat{b}}{2f^6}\(s^2+t^2+u^2\)+ \frac{3}{f^8}\(\frac{3\Delta a}{2}-\hat{b}^2\)s t u +\cdots\ ,
\end{align}
where $s,t,u$ are the usual Mandelstam variables. This 4-point amplitude suggests that there should be a dispersive sum-rule for the parameter $\hat{b}$ which is identical to the sum-rule for the 4D $a$-theorem of \cite{Komargodski:2011vj} (see \cite{Elvang:2012st}). This sum-rule would at first seem to imply that $\hat{b}$ is strictly positive, which is in tension with the analysis of section \ref{sec:example}.
Here we propose a possible resolution based on the large $s$ behavior of these scattering amplitudes.

We begin by carefully analyzing the derivation of such a sum-rule. The sum-rule follows from a contour integral
\be
\frac{1}{2\pi i}\oint ds \frac{\cA_4(s,t)}{s^3}=0\ ,
\ee
where the integral is performed over the same contour as  \cite{Elvang:2012st}. This can be expanded to write
\be\label{eq:4ptamp}
\frac{1}{2}\p_s^2 \cA_4(s,t)|_{s=0}=\frac{1}{\pi}\int_{s_*}^\infty ds \(\frac{1}{s^3}+\frac{1}{(s+t)^3}\)\mbox{Im}\ \cA_4(s,t)+\frac{1}{2\pi i}\int_\infty ds \frac{\cA_4(s,t)}{s^3}\ ,
\ee
where $s_*$ is some IR cut-off and the last integral is over the contour at infinity.
This last term is generally dropped in 4D, however, we keep it in 6D since
as we explain next, it can potentially be important.
For a possible sum-rule, we proceed by taking the forward limit $t=0$
\be
\frac{\hat{b}}{2f^6}=\frac{2}{\pi} \int_{s_*}^\infty ds \frac{\mbox{Im}\ \cA_4(s,0)}{s^3}+\frac{1}{2\pi i}\int_\infty ds \frac{\cA_4(s,0)}{s^3}\ .
\ee
Now from the optical theorem, we know that $\mbox{Im} \cA_4(s,0) =s \sigma(s)$ where $\sigma(s)$  is the total cross section for $\phi\phi$ scattering. Unitarity implies that $\sigma(s)>0$ for any interacting theory and hence the first term of the right hand side is strictly positive. Therefore, $\hat{b}=0$ necessarily requires
\be\label{eq:asymptotic}
\int_\infty ds \frac{\cA_4(s,0)}{s^3}=-4i \int_{s_*}^\infty ds \frac{\sigma(s)}{s^2}
\ee
The above relation can only be satisfied if
\be\label{eq:asym}
\cA_4(s\rightarrow \infty,0) \gtrsim s^2\ .
\ee

At first sight, the condition (\ref{eq:asym}) seems to be violating the Froissart bound which in 6D imposes \cite{CHAICHIAN1992151}
\be
|\cA_4(s\rightarrow \infty,0)|\le (\text{const.}) s (\log s)^{4}\ .
\ee
However, the Froissart bound does not apply to the 4-point scattering of Goldstone modes, since there is no mass gap.

Even though the Froissart bound is not applicable, we can still say a few things
about asymptotic limits of dilaton amplitudes. In particular, the limit
$s\rightarrow \infty$ of $\cA_4(s,0)$ is dictated by the coupling of the dilaton to operators of the UV CFT \cite{Komargodski:2011xv,Luty:2012ww}. In general, the dilaton can couple to any relevant operator (that acquires a VEV) of the UV CFT. So, consider a generic relevant operator $\cO_{\rm UV}$ of dimension $\Delta_{\rm UV}$ of $\CFTUV$ which has a coupling with the dilaton
\be
\frac{\lambda}{f^{2n}}\int d^6x \phi^n \cO_{\rm UV}
\ee
where only $n=1,2,3$ can contribute to a four-point scattering amplitude of the dilaton. Note that $\lambda$ must have dimension $6-\Delta_{\rm UV}$. Moreover, the factor of $f$ follows from the fact that the actual coupling should be between  $\CFTUV$ and the $\tau$, $\beta_a$ fields. The $\tau$, $\beta_a$ fields have expansion in terms of $\frac{\phi}{f^2}$. Now the contribution of all possible such couplings to the amplitude $\cA_4(s,0)$ is fixed from dimensional analysis
\be
\cA_4(s\rightarrow \infty,0) \sim \frac{\lambda^N}{f^8}s^{\frac{N}{2}\Delta_{\rm UV}-3(N-1)}\
\ee
where $N$ is an integer, with $N\in \{2,3,4\}$. For example, $N=4$ comes from a coupling $\int \phi \cO_{\rm UV}$. Thus, we can derive an upper bound on the asymptotic amplitude
\be
\cA_4(s\rightarrow \infty,0) <(\text{const.}) s^3
\ee
which is consistent with (\ref{eq:asym}). Furthermore, (\ref{eq:asym}) necessarily requires at least one
\be
\Delta_{\rm UV} \ge 5\ .
\ee
Note that this bound is a direct consequence of $\hat{b}=0$.

Let us also note that a similar analysis of the coupling of the dilaton to operators of the UV CFT in 4D leads to $\cA_4(s\rightarrow \infty,0) <(\text{const.}) s^2$ \cite{Komargodski:2011xv,Luty:2012ww}. Hence in 4D, one can safely ignore the contribution from infinity to the dispersion relation (\ref{eq:4ptamp}). This explains why $\Delta a$ is strictly positive in 4D.

Of course, the above discussion immediately begs the question: If there are important modifications to dilaton scattering at large $s$,
where do they come from? In fact, there is a natural culprit for what can potentially modify the
structure of dilaton scattering at large $s$ in $D>4$ SCFTs:\ the appearance
of the effective strings themselves! Recall that in a mixed branch flow where we have the scales $f_t$ and $f_H$, the case of a Higgs branch
flow can be understood as coming from the regime where $f_t^{2}<s < f_H^2$.
As such, we are directly sensitive to the stringy excitations on the worldsheet of these effective
strings. We emphasize that this is qualitatively different from what happens in the case of 6D $\mathcal{N} = (1,0)$ tensor branch flows, and their counterparts in 6D $\mathcal{N} = (2,0)$ SCFTs, since in those cases we would have required $s < f_t^2$.
What we need to know is whether there are any effective string
excitations which could potentially modify the large $s$ behavior of
4-dilaton scattering amplitudes.

First of all, these effective strings are closed strings. From the structure
of 6D\ supersymmetry, we can already make out the appearance of an
excitation associated with the anti-chiral 2-form of the tensor multiplet.
There is, however, no massless graviton excitation
(since 6D gravity is decoupled). Rather, we expect there
to be (in 4D\ terms) a \textquotedblleft massive spin 2
excitation\textquotedblright, i.e. its Regge intercept is close to two. This
is \textit{precisely} the large $s$ scaling required to modify the profile of
4-dilaton scattering, since at Regge intercept precisely two we would have
an amplitude which grows as $\mathcal{M}_{2\rightarrow2}\sim s^{2}$. In the case of an intercept near two, we
expect a more general scaling of the form $s^{2+\varepsilon}$, where a priori,
$\varepsilon$ could be either positive or negative. On physical grounds, we expect
that $\epsilon\rightarrow 0$ as we take the strict $f_t\rightarrow 0$ limit.

Let us sketch some basic elements of these effective string theories.
While it is certainly challenging to extract much precise information,
some properties such as the anomalies and
elliptic genera of the worldsheet CFT are available (see e.g. \cite{Minahan:1998vr, Gadde:2015tra,
Shimizu:2016lbw, DelZotto:2016pvm, Apruzzi:2016nfr}). Observe that there are four transverse
directions to the worldvolume of the string, and with it, four massless
scalars. The creation operators for these scalars on the worldsheet allow us
to build (in the obvious notation)\ the standard states of the form
$\alpha_{\mu}\widetilde{\alpha}_{\nu}\left\vert k\right\rangle $. In
particular, in the indices $\mu$ and $\nu$ we can take the trace (for the
dilaton), anti-symmetrize and project onto chiral and anti-chiral two-forms,
or symmetrize (for the massive spin 2 excitation). This is in accord with the
fact that the massless tensor multiplet contains a scalar and an anti-chiral
two-form, while the 6D massless graviton multiplet contains a graviton and a
chiral two-form. That being said, extracting the precise Regge trajectories
for massive string excitations in this case appears quite challenging, and is a topic
we leave for future investigation.\footnote{We thank S. Hellerman for helpful correspondence.}

What all of this points to is that at least in 6D (as well as in all known
$D>4$ SCFTs), the large $s$ behavior of dilaton scattering can receive
potentially important corrections. These corrections can be neglected in the
case of tensor branch deformations, but in the case of both Higgs branch and
mixed branch flows for 6D\ SCFTs, the situation is far less clear. In fact,
all indications thus far point to the \textquotedblleft seemingly
naive\textquotedblright\ answer that $\hat{b}_{\text{Higgs}}=0$, though it
may indeed still be the case that $\hat{b}_{\text{Higgs}}>0$ (even though
there is as yet no direct evidence for such an assertion).

An additional final comment is that these considerations mainly impact
4-point dilaton scattering amplitudes, but do not appear to affect
scattering for $n$-point amplitudes for $n\geq6$ since in the latter case the
dimensionality of the phase space is large enough to bypass such subtleties.



\section{Conclusions \label{sec:CONC}}

Six-dimensional conformal field theories exhibit a number of features distinct from
$D \leq 4$ CFTs. That being said, some aspects of these systems can still be constrained using
techniques from effective field theory. In this paper we have investigated the structure of conformal symmetry breaking
in such theories, when it is accompanied by the breaking of a continuous global symmetry. The resulting
dilaton-axion effective action exhibits some important constraints, and relates different higher-dimension operators.
In particular, we have used this formalism to study the structure of some higher-derivative interaction terms
in 6D SCFTs, including the cases of tensor branch flows, Higgs branch flows, and mixed branch flows. In the case of tensor branch flows,
this analysis provides a way to determine the precise numerical relation between
the Euler conformal anomaly to the coefficient of $\phi^2 \Box^2 \phi^2$, with $\phi$ the dilaton for
spontaneous conformal symmetry breaking.
In the case of mixed branch flows we also determined the precise contributions to the coefficient of $\phi^2 \Box^2 \phi^2$, as a function of the scales $f_t$ and $f_H$ associated with tensor branch and Higgs branch deformations. We also presented some examples
of 6D SCFTs illustrating that actually generating contributions to $\phi^2 \Box^2 \phi^2$ on the Higgs branch appears to be surprisingly challenging. This led us to conjecture that in stark contrast to all $D \leq 4$ CFTs, it may be possible to have an interacting dilaton effective field theory, even when the term $\phi^2 \Box^2 \phi^2$ vanishes. In the remainder of this section we turn to some further avenues of investigation.

As we have already mentioned, one of the surprising aspects of our analysis is the difficulties which appear in generating an explicit
contribution to $\phi^2 \Box^2 \phi^2$ in 6D Higgs branch flows. The standard lore, which is very well supported in $D \leq 4$ CFTs, is that the coefficient of this term must be strictly positive in order to have an interacting effective field theory. There is an important caveat to such dispersion relation arguments, since it presupposes a particular falloff for dilaton scattering at large $s$.
Indeed, compared with the case of $D \leq 4$ CFTs, in 6D SCFTs, effective strings play a crucial role in even realizing a fixed point in the first place. Incorporating the effects of these effective strings is of course rather challenging, but it points to a general issue which should be properly addressed rather than just ``taken for granted''. One way to settle the impact on dilaton scattering would be to study the Regge intercept of the corresponding effective strings. From general considerations, we know that this intercept is close to two, and so it has a chance to make an important contribution to the large $s$ behavior of dilaton scattering. Another way to address this issue would be to explicitly exclude the presence of supersymmetry preserving deformations which could generate such a $\phi^2 \Box^2 \phi^2$ interaction term in a Higgs branch flow, perhaps generalizing the analysis presented in reference \cite{Cordova:2015fha}.

At a general level, it would also be interesting to study the full structure
of the $40+40$ dilaton-Weyl multiplet for 6D $\mathcal{N} = (1,0)$ theories,
and the associated dilaton effective action in this case \cite{Bergshoeff:1985mz}.
This would likely provide additional information on the
structure of general flows in 6D\ SCFTs.

In this paper we have focused on the structure of
dilaton scattering amplitudes. Based on general supersymmetric considerations,
it is natural to expect that at least for 6D SCFTs, these higher-derivative interaction terms are
closely related to \textquotedblleft topological\textquotedblright terms.
It would be interesting to study such structures, and their relation
to anomalies of the associated field theories.

Finally, in references \cite{Baume:2020ure, Heckman:2020otd} it was noted that there are also specific
nearly protected operator subsectors in 6D\ SCFTs which resemble a
one-dimensional quantum integrable spin chain. To further probe the structure
of this sector, it would be quite interesting to consider the structure of
dilaton scattering off of backgrounds with a finite chemical potential for
such states. Similar considerations apply to the analysis of dilaton scattering off of supersymmetric defects of the sort
considered for example in \cite{DelZotto:2015isa, Wang:2020xkc}.


\section*{Acknowledgments}

We thank T. Rudelius for collaboration at an early stage of this work. We also thank
C.P. Herzog and D.S. Park for helpful discussions, and S. Hellerman and M. Ro\v{c}ek for helpful correspondence.
We thank Z. Komargodski and S. Theisen for comments on an earlier draft. We especially thank
C. C\'{o}rdova, T.T. Dumitrescu and K. Intriligator for comments on an earlier draft.
The work of JJH is supported by a University Research Foundation grant at the University of Pennsylvania and
DOE (HEP) Award DE-SC0021484. The work of SK is supported by the Simons
Collaboration Grant on the Non-Perturbative Bootstrap.


\appendix

\section{The Dilaton-Axion Effective Action}\label{app:DILAX}
\subsection{$SU(2)$ Global Symmetry}

In this Appendix we summarize for the sake of completeness some additional details of the dilaton-axion effective action.
As discussed in \cite{Kundu:2020bdn}, the following 4-derivative invariants can be constructed for $SU(2)$:
\begin{align}\label{def_W}
&W_1=\Tr \(\(\hat{\nabla}^\mu \hat{A}_\mu\)\( \hat{\nabla}^\nu \hat{A}_\nu\)\) \ , \nonumber\\
&W_2=\hat{g}^{\mu\nu}\Tr \(\hat{A}_\mu \hat{\Box}\hat{A}_\nu\)\nonumber \\
&W_3=\hat{R}^{\mu\nu}\Tr \(\hat{A}_\mu\hat{A}_\nu\)\ , \nonumber\\
&W_4=\hat{R}\hat{g}^{\mu\nu}\Tr\(\hat{A}_\mu\hat{A}_\nu\)\ ,  \nonumber\\
&W_5=\hat{g}^{\mu\nu}\Tr\(\hat{A}_\mu\hat{A}_\rho \hat{\nabla}^\rho \hat{A}_\nu\)\ , \nonumber\\ &W_6=\hat{g}^{\mu\nu}\hat{g}^{\rho\sigma}\Tr\(\hat{A}_\mu\hat{A}_\nu\hat{A}_\rho\hat{A}_\sigma\)\ , \nonumber\\
&W_7=\hat{g}^{\mu\nu}\hat{g}^{\rho\sigma}\Tr\(\hat{A}_\mu\hat{A}_\nu\)\Tr\(\hat{A}_\rho\hat{A}_\sigma\)\ , \nonumber\\
&W_8=\hat{g}^{\mu\nu}\hat{g}^{\rho\sigma}\Tr\(\hat{A}_\mu\hat{A}_\rho\)\Tr\(\hat{A}_\nu\hat{A}_\sigma\)\ ,
\end{align}
where, $\hat{R}$, $\hat{R}^{\mu\nu}$, and $\hat{\D}_{\mu}$ are computed using the Weyl-invariant metric (\ref{metric0}). Hence, at the 4-derivative order we can write:
\be
\int d^6x \sqrt{\hat{g}} \cL_{SU(2)}[\hat{g}_{\mu\nu},\hat{A}_\mu]=\int d^6x \sqrt{\hat{g}} \sum_{I=1}^8 W_I\ .
\ee
The invariants $W_I$ in the flat space limit with no background gauge field can be further simplified by using the equations of motion. In fact, not all the $W_I$ are independent.  In particular, inside the integral, at the four-field level we obtain:
\begin{align}\label{def_W1}
&W_1=0\ ,\qquad W_2=W_3=-\frac{1}{16f^8}\(3\xi^2 \Box^2 \xi^2+2\phi \xi_a \Box^2 \phi \xi_a+\xi^2 \Box^2 \phi^2\)\  ,   \\
&W_4=-\frac{5}{f^8}\xi^2 \Box^2 \xi^2\ , \qquad W_5=\frac{1}{8f^8}\(\delta_{ab}\delta_{cd}-\delta_{ac}\delta_{bd}\) \( \xi_a   \xi_b\)\Box^2 \(\xi_c   \xi_d\)\ ,\nonumber\\
&W_6=\frac{1}{32f^8}\xi^2\Box^2 \xi^2\ , \qquad W_7=\frac{1}{16f^8}\xi^2\Box^2 \xi^2\ , \qquad W_8=\frac{1}{16f^8}\( \xi_a   \xi_b\)\Box^2 \( \xi_a   \xi_b\)\ ,\nonumber
\end{align}
where $\xi^2=\xi_a\xi_a$.

The rest of the analysis is almost identical to that of \cite{Kundu:2020bdn}. Putting everything together, we find that the low energy effective action has the form of equation (\ref{eq:eff}), where
\begin{align}\label{app:axions}
\cL_{\rm axion}[\xi]=\frac{B_0}{f^4}\epsilon_{abc}\epsilon_{ab'c'}\xi_b \xi_{b'}\Box \(\xi_c \xi_{c'}\) +\frac{B_1}{f^6} \xi^2\Box^2 \xi^2 + \frac{B_2}{f^6}\( \xi_a   \xi_b\)\Box^2 \( \xi_a   \xi_b\) +\cO\(\xi^6, \p^6\)\ ,
\end{align}
and $B_0,B_1,B_2$ are dimensionless coefficients. Note that the two-derivative interactions in $\cL_{\rm axion}[\xi]$ come from the second term of (\ref{eq:kinetic}). Similarly, the mixed interactions are given by
\be\label{app:mixed}
\cL_{\rm mixed}[\phi,\xi]=\frac{\tilde{B}_1}{f^6}\phi \xi_a \Box^2 \( \phi \xi_a\)+ \frac{\tilde{B}_2}{f^6}\xi^2 \Box^2 \phi^2+\cO\(\xi^2\phi^3, \p^6\)\ ,
\ee
where $\tilde{B}_1,\tilde{B}_2$ are dimensionless coefficients. Finally, the most important part of the effective action $\cL_{\rm dilaton}[\phi]$ is given by equation (\ref{eq:dilaton}). Of course, for supersymmetric flows there will be massless fermionic degrees of freedom which can contribute to amplitudes of bosonic degrees of freedom via loop diagrams.

\subsection{Other Global Symmetries}

It is a straightforward exercise to generalize to RG flows between two conformal fixed points in 6D in which
$\CFTUV$ has some global symmetry $G$, where $G$ is an arbitrary compact Lie group. The conformal symmetry of $\CFTUV$ is broken either spontaneously or explicitly which triggers the RG flow that breaks the global symmetry $G$ to  some subgroup $H$. All such RG flows can be described by an effective field theory of Goldstone  bosons of spontaneously broken conformal and global symmetries. The effective field theory consists of a {\it dilaton} $\phi$ from the broken conformal symmetry and $N=\mbox{dim}\ G/H$ {\it axions} $\xi_a$  arising from the broken global symmetry.  We assume that  $G$ is a direct product of a finite number of simple Lie groups and $U(1)$ factors:
\be
G=\prod_I G_I
\ee
and extend the  general framework of \cite{Kundu:2020bdn} to 6D. The low energy effective action has the form of equation (\ref{eq:eff}), where  the dilaton effective action $\cL_{\rm dilaton}[\phi]$ is still given by equation (\ref{eq:dilaton}). The axion effective action is now
\begin{align}
\cL_{\rm axion}[\xi]=\frac{B_0}{f^4}f_{abc}f_{ab'c'}\xi_b \xi_{b'}\Box \(\xi_c \xi_{c'}\) +\frac{B_{abcd}}{f^6} \( \xi_a   \xi_b\)\Box^2 \( \xi_c   \xi_d\) +\cO\(\xi^6, \p^6\)\ ,
\end{align}
where axions $\xi_a$ with $a \in {1, 2,\cdots , N}$ belong in some large reducible representation with structure constants $f_{abc}$. Similarly, the mixed interactions are given by
\be
\cL_{\rm mixed}[\phi,\xi]=\frac{\tilde{B}^{(1)}_{ab}}{f^6}\phi \xi_a \Box^2 \( \phi \xi_b\)+ \frac{\tilde{B}^{(2)}_{ab}}{f^6}\xi_a \xi_b \Box^2 \phi^2+\cO\(\xi^2\phi^3, \p^6\)\ .
\ee
Therefore, breaking of any global symmetry will not affect dilaton amplitudes as obtained from $\cL_{\rm dilaton}[\phi]$ at the 6-field 6-derivative order.



\bibliographystyle{utphys}
\bibliography{RGBib}

\end{document}